\documentclass[11pt]{article}							

\usepackage[utf8]{inputenc}							
\usepackage[english]{babel}	
\usepackage[babel=true]{csquotes}
\usepackage{color, colortbl}
\usepackage{float}							
\usepackage{graphicx}							
\usepackage{amsmath,amssymb}							
\usepackage{color}							
\usepackage{array}							
\usepackage{amsfonts}							
\usepackage{dsfont}							
\usepackage{hyperref}
\usepackage{multirow}
\usepackage{hyperref}
\usepackage{natbib}
\usepackage{amsmath}
\usepackage{graphicx,psfrag,epsf}
\usepackage{bm}
\usepackage{caption}
\usepackage{color}	
\usepackage{float}
\usepackage{multirow}
\usepackage{dsfont}
\usepackage[flushleft]{threeparttable}
\usepackage{longtable}

\definecolor{amber}{rgb}{1.0, 0.49, 0.0}

\begin{document}							

\begin{center}							
{\Large							
   {\sc  Dynamic modelling of Multivariate Dimensions and Their Temporal Relationships using Latent Processes: Application to Alzheimer's Disease							
   }							
}							
							
\bigskip							
Bachirou O. Taddé $^{1,^\star}$, Hélène Jacqmin-Gadda$^{1}$, Jean-Fran\c ois Dartigues$^{1}$, Daniel Commenges$^{1}$, Cécile Proust-Lima $^{1}$  for the Alzheimer’s Disease Neuroimaging Initiative$^{\dagger}$

\vspace{1mm}
				
\bigskip							
							
{\it							
							
$^{1}$ INSERM, UMR1219, Univ. Bordeaux, F33000 Bordeaux, France\\							
$^\star$ oladedji-bachirou.tadde@u-bordeaux.fr\\ 
$^{\dagger}$Data used in preparation of this article were obtained from the Alzheimer’s Disease
Neuroimaging Initiative (ADNI) database (adni.loni.usc.edu). As such, the investigators
within the ADNI contributed to the design and implementation of ADNI and/or provided data
but did not participate in analysis or writing of this report. A complete listing of ADNI
investigators can be found at: http://adni.loni.usc.edu/wp-content/uploads\/how\_to\_apply/ADNI\_Acknowledgement\_List.pdf	
}							
\end{center}							
\bigskip							
							
%--------------------------------------------------------------------------							

\begin{center}							
\bf{Abstract}							
\end{center}  							
Alzheimer's disease gradually affects several components including the cerebral dimension with brain atrophies, the cognitive dimension with a decline in various functions and the functional dimension with impairment in the daily living activities. Understanding how such dimensions interconnect is crucial for AD research. However it requires to simultaneously capture the dynamic and multidimensional aspects, and to explore temporal relationships between dimensions. We propose an original dynamic structural model that accounts for all these features. The model defines dimensions as latent processes and combines a multivariate linear mixed model and a system of difference equations to model trajectories and temporal relationships between latent processes in finely discrete time. Dimensions are simultaneously related to their observed (possibly multivariate) markers through nonlinear equations of observation. Parameters are estimated in the maximum likelihood framework enjoying a closed form for the likelihood. We demonstrate in a simulation study that this dynamic model in discrete time benefits the same causal interpretation of temporal relationships as models defined in continuous time as long as the discretization step remains small. The model is then applied to the data of the Alzheimer's Disease Neuroimaging Initiative. Three longitudinal dimensions (cerebral anatomy, cognitive ability and functional autonomy) measured by 6 markers are analyzed and their temporal structure is contrasted between different clinical stages of Alzheimer's disease. 							

\smallskip							


%--------------------------------------------------------------------------							
\newpage

\section{Introduction}\label{s:intro}

Dementia is a general syndrome characterized by a long term and gradual decrease in the ability to think and remember with consequences on the person's daily functioning. It represents a pressing public health problem with an estimated worldwide prevalence of 46.8 millions cases and health care burden with US $\$$ 817.9 billions in 2015  \citep{wimo2017worldwide}. %There is currently no cure although many novel compounds are under development. 
Alzheimer disease (AD), the most common form of dementia (60\% to 80\% of the cases, \cite{reitz2014alzheimer}), gradually affects multiple components long before clinical diagnosis with brain atrophies, cognitive decline in various functions (memory, language, orientation in space and time, etc.) and loss of autonomy in daily living activities. 

Although essential for the assessment of compounds and more generally for AD research, knowledge of the natural history of AD and its progression is still imprecise. Theoretical schemes have highlighted its expected dynamic and multidimensional aspects. For instance, \cite{jack2013tracking} hypothesized a sequence with the accumulation of proteins in the brain (amyloid-$\beta$ and $\tau$ proteins), the atrophy of brain regions (e.g., hippocampus) and then clinical manifestations with cognitive and functional declines on which dementia diagnosis currently relies. However, this theory is hard to translate into a statistical model because it requires combining multidimensional and dynamic aspects, and exploring temporal relationships between dimensions.	

As in other medical contexts, the dynamic aspect of AD has been mostly apprehended using mixed models in which one dimension is analyzed according to time and covariates \citep{amieva2008prodromal,donohue_estimating_2014}. This approach naturally handles dynamic processes by modeling in continuous time the process underlying the repeated discrete observations. To emphasize the distinction between the true process of interest from its noisy observations, mixed models can be split into a structural model which analyzes the latent process over time, and equations of observations which simultaneously link the unobserved quantity to the noisy observed outcomes at each visit \citep{proust2013analysis}. With such split, mixed models can treat one or more markers indiscriminately, as long as they measure the same latent quantity. This is particularly useful for dimensions such as cognition measured by a battery of neuropsychological tests or cerebral anatomy measured by regional volumes.

To understand how dimensions are inter-related in AD, some have explored pre-determined relationships by examining change over time of one biomarker %(e.g., cognitive test) 
according to another one %(e.g., hippocampal initial volume) 
and assumed the latter was observed without measurement error (e.g., \cite{landau20111207, han2012beta}). This approach quantifies temporal relationships but it relies on a specific \emph{a priori} determined sequence and does not consider all biomarkers as continuous processes. Others used bivariate mixed models to dynamically model two dimensions and account for their correlation through correlated random effects (e.g., \cite{mungas2005longitudinal,robitaille2012multivariate}). However such models remain descriptive; they account for the correlation between the markers but do not allow to distinguish the influence of each marker on the other.

Temporal asymmetric relationships between processes have been mainly apprehended with Dynamic Bayesian Networks (DBN), Dynamic Structural Equations Modelling (DSEM) and Cross-Lagged Models (CLM) \citep{song2009time, hamaker2015critique}. These approaches extend the concept of Directed Acyclic Graphs (DAG) \citep{greenland_causal_2000} to longitudinal data by modelling temporal relationships between successive states of a network of processes. Although they account for the longitudinal data structure and the measurement sequence, these methods have two main drawbacks that this contribution intends to circumvent. First, the definition of the model most often depends on the discrete visit process, so that time associations are limited to those between successive observed visits \citep{hamaker2015critique, kuiper2018drawing,voelkle_role_2018}. 
% Although they account for the longitudinal data structure and the measurement sequence, they do not explicitly account for the time elapsed between two occasions \citep{hamaker2015critique, kuiper2018drawing}. 
This can lead to biased results in case of unequal time intervals. Even when observations are equally-spaced (generally in grossly discrete time), or when the time elapsed between visits is taken into account through time-dependent parameters, the estimated lagged effects are still specific to the discrete time intervals used in the study and spurious causal temporal associations might appear as recently demonstrated \citep{aalen2016can}. The second and main drawback of these methods is that they quantify the association between successive levels of the processes while we are mainly interested by the influence of each process on the subsequent change of other processes. Yet a dynamic view of causality would seek local dependence structures linking the network of processes to its infinitesimal subsequent change over time \citep{aalen2007Whatcan,didelez2008,commenges_general_2009} in order to retrieve the mechanism which explains how the system changes as time changes \citep{voelkle_role_2018}. Local dependence structures can be naturally investigated with mechanistic models which relate a system of processes over time using differential equations. Proposed notably in HIV studies \citep{prague_dynamic_2017}, they allow retrieving causal associations between disease components. Yet mechanistic models are numerically very demanding so that their application to complex diseases such as AD is compromised. In addition they require precise biological knowledge which lacks for AD.

Our objective is thus to propose a statistical model that simultaneously describes the dynamics of multiple dimensions involved in AD and assesses their temporal relationships similarly as in a mechanistic model but with much less numerical complexity. We consider a system of latent dimensions possibly observed through one or several longitudinal markers. We define a set of difference equations to model the change over a discretized time of the system according to its previous state. Yet in contrast with other methods (DBN, DSEM, CLM), this discretization step is disconnected from the observation process and can be finely chosen. As discretization might still distort the causal interpretations of temporal relationships compared to a model in continuous time, we specifically evaluate the impact of discretization on the temporal influence structure in a simulation study. The methodology is applied to the Alzheimer's Disease Neuroimaging Initiative database to explore the temporal structure between cerebral, cognitive and functional dimensions at different stages of AD.	

\section{Motivating Data}\label{data} 														
Data were obtained from the Alzheimer's Disease Neuroimaging Initiative (ADNI) database \\(adni.loni.usc.edu). The ADNI was launched in 2003 with the primary goal to test whether serial magnetic resonance imaging (MRI), positron emission tomography (PET), other biological markers, and clinical and		
neuropsychological assessment could be combined to measure the progression of mild			
cognitive impairment (MCI) and early Alzheimer's disease (AD). 														
We focused on the ADNI 1 phase \citep{mueller2005alzheimer}, a multisite observational study that included N$\approx$800 individuals aged 55-90 enrolled in three stages of progression to AD: normal aging (CN, N$\approx$200), mild cognitive impairment (MCI, N$\approx$400), and diagnosed Alzheimer's Disease (dAD, N$\approx$200). Participants were followed-up every 6 months up to 3 years for CN and MCI groups and up to 2 years for the dAD group (except at month 18 for dAD and CN, and at month 30 for CN and MCI). At each visit, information was collected, including for the present work, volumes of brain regions measured by MRI, a battery of 19 cognitive scores and a functional assessment.

\section{Methodology}\label{Methodology}														
\subsection{Structural model for the system of latent processes}\label{Structural_model}														
Consider $D$ latent processes $\boldsymbol{\Lambda_i}(t)_{t \geq 0}$ (with $ \boldsymbol{\Lambda_i}(t) = (\Lambda_{i}^d(t))^{\top}_{d=1,...,D}$) representing  a system of $D$ dimensions (e.g. $D=3$ for cerebral anatomy, cognitive ability and functional autonomy dimensions in AD) for individual $i$ with $i=1, \dots, N$. 
%As the methodology relies on a system of difference equations, we discretize the time with constant discretization step $\delta$. 
We assume $\boldsymbol{\Lambda_i}(t)_{t \geq 0}$ is defined at discrete times $t_j =  j \times \delta$ with $j \in \tau = \{0, 1, \dots, J\}$, and $\delta$ a constant discretization step. 
%Pourquoi pas delta x j, j=1,...,J  ??
Let us denote $\Delta\boldsymbol{\Lambda_i}(t_j + \delta)= \boldsymbol{\Lambda_i}(t_j + \delta) - \boldsymbol{\Lambda_i}(t_j)$ the change of the system between two successive times and $\frac{\Delta\boldsymbol{\Lambda_i}(t_j + \delta)}{\delta}$ the rate of change of the system.  		

To make explicit the modelling of the temporal relationships, we split the structural model into two multivariate linear mixed submodels for: (i) the level of the processes at baseline $\boldsymbol{\Lambda_i}(0)$ and (ii) the rate of change of the system over time $\frac{\Delta\boldsymbol{\Lambda_i}(t_j + \delta)}{\delta}$ using difference equations: 											
\begin{equation}\label{struct_model_1}														
\left\{														
\begin{array}{rl}														
\boldsymbol{\Lambda_i}(0) = & \boldsymbol{X_i^0} \boldsymbol{\beta} + \boldsymbol{u_i}\\														
\frac{\Delta\boldsymbol{\Lambda_i}(t_j + \delta)}{\delta} = &  \boldsymbol{X_i}(t_j + \delta)\boldsymbol{\gamma} + \boldsymbol{Z_i}(t_j +\delta)\boldsymbol{v_i} + \boldsymbol{A_{i,\delta}}(t_j) \boldsymbol{\Lambda_i}(t_j)%,~~~\forall ~t_j \geq 0  ,														
\end{array}														
\right.														
\end{equation}														
where $\boldsymbol{X_i^0}$ is the $D \times p_0$-matrix of covariates associated with the $p_0$-vector of fixed effects $\boldsymbol{\beta}$, and $\boldsymbol{u_i}$ is the $D$-vector of individual random intercepts $u_i^d$ in the initial system $\boldsymbol{\Lambda_i}(0)$. The $(D\times p)$-matrix $\boldsymbol{X_i}$ and the $(D\times q)$-matrix $\boldsymbol{Z_i}$ include time-dependent covariates associated with the $p$-vector of fixed effects $\boldsymbol{\gamma}$ and the $q$-vector $(q=\sum_{d=1}^{D}q_d )$
%% and the $\q=\sum_{d=1}^{D}q_d$-vector 
of individual random effects $\boldsymbol{v_i}=({v_i^d}^\top)_{d=1,...,D}$, respectively. $\boldsymbol{A_{i,\delta}}$ is the $D\times D$-matrix of temporal influences.

For each process $\Lambda_i^d$, the $(q_d +1)$ vector of individual random effects ${(u_i^d, {v_i^d}^\top)}^{\top}$ is assumed to have a multivariate normal distribution with variance-covariance matrix 
$\left ( \begin{array}{cc} {b_u^d}^2 & \boldsymbol{B_{uv}^d} \\ {\boldsymbol{B_{uv}^d}}^\top & \boldsymbol{B_v^d} \end{array} \right ) $
where $\boldsymbol{B_{uv}^{d}}$ and $\boldsymbol{B_v^d}$ are unstructured.
%\cpl{As explained in next section, to reach identifiability,} the variances of the $u_i^d$s are constrained to 1 so that the D processes have the same magnitude at baseline. 
We assume that random effects may be correlated within
each process $d$ (to take into account inter-individual variability in each process trajectory)
and between processes at baseline (to take into account the possible
within-individual correlation between processes due to anterior dependencies). 
%\bt{Variance-covariance matrix of all random effects of the structural model is suitable used to account for intra-processes correlations by correlating random effects within each process $\Lambda_i^d$  and to account for inter-processes correlations at baseline due to anterior dependencies by correlating random effects on the initial levels of processes.} 
Other correlations between random effects are kept null to ensure that the temporal relationships are entirely captured by matrix $\boldsymbol{A_{i,\delta}}$. Consequently the entire $(D+q)$-vector of individual random effects $\boldsymbol{w_i}=(\boldsymbol{u_i}^{\top},\boldsymbol{v_i}^{\top})^{\top}$ has a multivariate normal distribution,	
\begin{equation*}														
\boldsymbol{w}_i \sim 														
\mathcal{N}\left( \left(														
\begin{array}{r}														
\boldsymbol{0} \\														
\boldsymbol{0}														
\end{array}														
\right), \boldsymbol{B}=\left(\begin{array}{cc}														
							\boldsymbol{B_u} & \boldsymbol{B_{uv}} \\							
 							   \boldsymbol{B_{uv}}^{\top} & \boldsymbol{B_v}							
						\end{array}								
				   \right)										
		   \right).												
\end{equation*}														
with $\boldsymbol{B_u}$ the variance covariance matrix of $\boldsymbol{u_i}$, $\boldsymbol{B_v}$ the D-block diagonal matrix with $d^\text{th}$  block $\boldsymbol{B_v^d}$, and $\boldsymbol{B_{uv}}$ the $D \times q$ matrix with $d^\text{th}$ row $\left(\begin{array}{ccc} \boldsymbol{O}_{\sum_{l=1}^{d-1}q_l} ,& \boldsymbol{B_{uv}^d} ,& \boldsymbol{O}_{\sum_{l=d+1}^{D}q_l} \end{array}\right)$ where $\boldsymbol{O}_x$ is the $x$-row vector of zeros.  														
In the estimation process, $\boldsymbol{B}$ is replaced by its Cholesky decomposition: $\boldsymbol{B} = \boldsymbol{L L}^\top$, where $\boldsymbol{L}$ is a $ (D+q) \times (D+q) $ lower triangular matrix. 

As in any latent variable model, the dimensions of the latent processes have to be defined to reach identifiability. Since we do not want to constrain the measurement models (in Section \ref{Measurement_Model}), we chose to standardize the latent processes at baseline by excluding intercepts from $\boldsymbol{X^0_i}$ (i.e., the processes have zero mean in the reference group at baseline) and fixing the variances of $\boldsymbol{u_i}$ at one ($b_u^d=1 ~ \forall d \in \{1,...,D\}$). 
%Constraints in the variance-covariance matrix of the random effects are handled by fixing corresponding parameters in $\boldsymbol{L}$.

The temporal influences between processes are modelled through the $D\times D$-matrix of time-dependent effects $\boldsymbol{A_{i,\delta}}(t_j)$:
\begin{equation*}														
\boldsymbol{A_{i,\delta}}(t_j)= \left( 					
\begin{matrix}														
a_{i,11}(t_j) & \ldots & a_{i,1d}(t_j) & \ldots & a_{i,1D}(t_j) \\		
\vdots & \ddots & \vdots & \ddots & \vdots\\					
a_{i,d1}(t_j) & \ldots & a_{i,dd}(t_j) & \ldots & a_{i,dD}(t_j) \\
\vdots & \ddots & \vdots & \ddots & \vdots\\						
a_{i,D1}(t_j) & \ldots & a_{i,Dd}(t_j) & \ldots & a_{i,DD}(t_j)			
\end{matrix}														
\right)														
\end{equation*}														
This matrix captures the directed temporal influences between latent processes at time $t_j$ and subsequent rates of change of latent processes between times $t_j$ and $t_j + \delta $. Specifically, coefficient $a_{i,dd'}(t_j)$ quantifies the temporal effect of process $d'$ at time $t_j$ on process $d$. Each effect can be modelled according to time/covariates through a linear regression $ a_{i,dd'}(t_j) = \boldsymbol{R}^{\top}_{i}(t_j)\boldsymbol{\alpha}_{dd'}$ where $\boldsymbol{R_i}(t_j)$ is a r-vector of time-dependent covariates associated with the r-vector of regression coefficients $\boldsymbol{\alpha_{dd'}}^{\top} = (\alpha^m_{dd^{'}})^{\top}_{m=0,(r-1)}$.														
														
When the discretization step is not too large, the temporal influences intend to
%\bt{a laisser, tu n'as pas encore démontrer}
have the same causal interpretations as those of a model in continuous time (see Simulation Study 2).

\subsection{Measurement Models of the longitudinal markers} \label{Measurement_Model}														
														
Consider K ($K\geq D$) continuous longitudinal markers $\boldsymbol{Y_{ij}} = (Y_{ijk})^{\top}_{k = 1, \dots, K}$ that have been measured for subject $i$ at $(n_i+1)$ continuous times $t_{ijk}^* \in [t_{j} , t_{j+1} )$ with $t_{j}$ the corresponding discretized time and $j \in \tau_i$, with $\tau_i$ being any subset of $\tau$ of length $(n_i+1)$. By considering $\tau_i \subset \tau$, we can consider observations sparser than the grid of discretized times.

%For simplicity, we consider for the moment that the discretization step used in the structural model is the same as in the observations.
% finalement je laisse, ça va être pénible sinon
Following \cite{proust2013analysis}, we assume that the latent process $\boldsymbol{\Lambda_i^d}$ is the underlying common factor of $K_d$ markers ($K = \sum\limits_{d=1}^{D}K_d$) and we note $\mathcal{K}^d$ the set of marker subscripts associated with latent process $\Lambda_i^d$. We assume that a marker measures only one latent process.														
														
The link between a marker and its underlying latent process is defined by a marker-specific measurement model. 														
If marker $Y_{ijk}$ is Gaussian, the measurement model is a linear equation:														
\begin{equation}\label{measurement_model_1}														
 \frac{Y_{ijk}-\eta_{0k} }{\eta_{1k} } = \tilde{Y}_{ijk} = \Lambda_i^d(t_{j}) + \tilde{\epsilon}_{ijk},  ~~~~~ \forall ~ k \in \mathcal{K}^d ~\text{and}~ \forall ~ j \in \tau_i,													
\end{equation}														
where the vector of transformation parameters $\boldsymbol{\eta_k} = (\eta_{0k},\eta_{1k})^\top$ is used to get the standardized form $\tilde{Y}_{ijk}$ of the marker and $\tilde{\epsilon}_{ijk}$ are independent Gaussian errors with variance $\sigma_{k}^2$. \\														
In the more general case of a continuous marker (possibly non Gaussian), one may consider a nonlinear observation equation:														
\begin{equation}\label{measurement_model_2}														
 H_k(Y_{ijk}; \boldsymbol{\eta_{k}}) = \tilde{Y}_{ijk} = \Lambda_i^d(t_j) + \tilde{\epsilon}_{ijk},  ~~~~~ \forall ~ k \in \mathcal{K}^d ~\text{and}~ \forall ~ j \in \tau_i,														
\end{equation}														
where the link transformation $H_k$ comes from a family of monotonic increasing and continuous functions parameterized with $\boldsymbol{\eta_{k}}$. Again $\tilde{Y}_{ijk}$ is the transformed marker and $\tilde{\epsilon}_{ijk}$ are independent Gaussian errors with variance $\sigma_{k}^2$. % pas exactement répét car definition au dessus, specifique au cas normal
The link transformation $H_k$ can be defined from a basis of I-splines (which are integrated M-splines \citep{Ramsay1988}) in association with positive coefficients, thus providing an increasing bijective flexible transformation. We used here a quadratic I-splines basis with $p_k$ internal knots, ${(\boldsymbol{\mathds{I}_m})}_{m = 1,p_k+3}$, so that 														
\begin{equation}\label{transformation}														
H_k(Y_{ijk}; \boldsymbol{\eta_{k}}) = \tilde{Y}_{ijk} = \eta_{0k} +  \displaystyle\sum\limits_{m=1}^{p_k+3}\eta_{mk}^2 \boldsymbol{\mathds{I}_m}(Y_{ijk}),														
\end{equation}														
with $(\eta_{mk})_{m=0,p_k+3}$ the vector of parameters of the transformation. Since we constrained the latent processes dimensions, $\eta_{k}$ does not need to be constrained to reach model identifiability.
														
In the following, we denote $\boldsymbol{\Sigma}=\text{diag}((\sigma_k^2)_{k=1,...,K})$ the diagonal variance matrix of the vector of errors $\boldsymbol{\tilde{\epsilon}_{ij}}=(\tilde{\epsilon}_{ijk})_{k=1,...,K}^{\top}$ and $\boldsymbol{\eta} = (\boldsymbol{\eta_{k}^{\top}})^{\top}_{\{1,\dots,K\}}$, the total vector of transformation parameters for the $K$ markers. The vector of transformed markers $\boldsymbol{\tilde{Y}_{ij}}$ %(transformation of the vector $\boldsymbol{Y_{ij}}$ with parameters $\boldsymbol{\eta}$) 
is mapped to the system of latent processes $\boldsymbol{\Lambda_i}(t_j)$ through a $K\times D$ matrix $\boldsymbol{P}$ with element $(k,d)$ equal to 1 if marker $k$ measures latent process $d$ and zero otherwise: 														
\begin{equation}\label{matrix_form_measurement_model_3}														
\boldsymbol{\tilde{Y}_{ij}}= \boldsymbol{P\Lambda_i}(t_j) + \boldsymbol{\epsilon_{ij}}														
\end{equation} 														
%For the sake of readability, we have introduced for now the observation models in the case where all the markers were observed at each occasion for subject $i$ in $\bt{\tau}$. 
In practice the observation process may include intermittent missing observations for a subset of markers or for all the markers at any occasion $j \in \tau_i $, so that $K_{ij}^{*} \leq K$ markers are actually observed at occasion $j$ for subject $i$. Following the mixed model theory, we assume that observations are missing at random and note $\boldsymbol{\tilde{Y}^*_{ij}}$ the transformations of the actual $K_{ij}^{*}$-vector of observed markers $\boldsymbol{Y^*}_{ij}$ at occasion $j$. 
The model linking the processes $\boldsymbol{\Lambda_i}(t_j)$ to the transformed markers $\boldsymbol{\tilde{Y}^*_{ij}}$ can be easily adapted to the presence of intermittent missing data by considering a $K_{ij}^{*} \times K $ observation matrix $\boldsymbol{M_{ij}}$ where element $(k^*,k)$ equals 1 if marker $k$ is the $k^{*\text{th}}$ observed marker at occasion $j$ and 0 if not for $k=1,...,K$ and $k^*=1,...,K_{ij}^{*}$:														
\begin{equation}\label{measurement_model_3}														
\boldsymbol{\tilde{Y}^*_{ij}}= \boldsymbol{M_{ij}P\Lambda_i}(t_j) + \boldsymbol{\epsilon^*_{ij}}														
\end{equation} 														
with $\boldsymbol{\epsilon^*}_{ij}$ the vector of independent Gaussian errors with covariance matrix $\boldsymbol{M_{ij}\Sigma M_{ij}}^{\top} $. 										
%\change{We assumes that observations are missing at random.}														
														
\subsection{Estimation by maximum likelihood}\label{Estimation}											
														
%The parameters are estimated in the  framework. 														
														
\subsubsection{Distribution of the Latent Processes and transformed observations}\label{Estimation_Distributions} 

Although the model was introduced through two submodels, the marginal distribution of latent processes (and by extension of the transformed observations) can be easily computed. By recurrence, the structural model \eqref{struct_model_1} can be rewritten:														

\begin{equation}\label{struct_model_2}														
\left\{														
\begin{array}{lll}		
\boldsymbol{\Lambda_i}(t_j)& = \boldsymbol{X^{(0)}_i} \boldsymbol{\beta} + \boldsymbol{u}_i ~~~~ & \text{if $j = 0$} \\														
\boldsymbol{\Lambda_i}(t_j)& = \boldsymbol{\tilde{A}_{i,\delta}}(0)\Big\{\boldsymbol{X^{(0)}_i} \boldsymbol{\beta} + \boldsymbol{u_i}\Big\} + ~\displaystyle \delta\Big\{\boldsymbol{X_i}(t_{ij})\boldsymbol{\gamma} + \boldsymbol{Z_i}(t_j)\boldsymbol{v_i}\Big\} & \text{if $j=1$}\\
\boldsymbol{\Lambda_i}(t_j)& =\displaystyle\prod_{l=0}^{j-1}\boldsymbol{\tilde{A}_{i,\delta}}(t_l)\Big\{\boldsymbol{X^{(0)}_i} \boldsymbol{\beta} + \boldsymbol{u_i}\Big\} + ~\displaystyle \delta\Big\{\boldsymbol{X_i}(t_j)\boldsymbol{\gamma} + \boldsymbol{Z_i}(t_j)\boldsymbol{v_i}\Big\} 				& \text{if $j > 1$}\\
& ~~~~~~+ ~\displaystyle \delta\sum\limits_{s=1}^{j-1}\displaystyle\prod_{l=s}^{j-1}\boldsymbol{\tilde{A}_{i,\delta}}(t_l) \Big\{\boldsymbol{X_i}(t_s)\boldsymbol{\gamma} + \boldsymbol{Z_i}(t_s)\boldsymbol{v_i}\Big\}					
\end{array}														
\right.														\end{equation}
														
where $t_j = j \times \delta$ for $ j \in \tau$ and $\boldsymbol{\tilde{A}_{i,\delta}}(t_j)= \boldsymbol{I}_D + \delta \boldsymbol{A_{i,\delta}}(t_j)$.														
														
By introducing $\boldsymbol{\Psi_{i,\delta}}(t_0,j,s)$ for $t_0 \geq 0 $ and $s \leq j$ so that 														
														
\begin{equation}\label{struct_model_3}														
\boldsymbol{\Psi_{i,\delta}}(t_0,j,s)=\left\{														
\begin{array}{ll}														
\boldsymbol{I}_D , ~~~~ & \text{if $s = j$}\\														
\displaystyle\prod_{l=s}^{j-1}\boldsymbol{\tilde{A}_{i,\delta}}( t_0 + t_l) ~~~~ & \text{if $s < j$}														
\end{array}														
\right.														
\end{equation} 														
														
Equation \eqref{struct_model_2} can be rewritten														
\begin{equation}\label{struct_model_4}														
\boldsymbol{\Lambda_i}(t_j)= \boldsymbol{\Psi_{i,\delta}}(0,j,0)\Big (\boldsymbol{X^0_i} \boldsymbol{\beta} + \boldsymbol{u_i}\Big ) + \displaystyle \Big[\delta \sum\limits_{s=1}^{j} \boldsymbol{\Psi_{i,\delta}}(0,j,s)\Big\{\boldsymbol{X_i}(t_s)\boldsymbol{\gamma} + \boldsymbol{Z_i}(t_s)\boldsymbol{v_i} \Big\}\Big]\mathds{1}_{ \{j > 0\}}														
\end{equation}				

From this equation, it can be seen that the structural model is a specific nonlinear mixed model depending on individual random effects $\boldsymbol{w_i}=(\boldsymbol{u_i}^{\top},\boldsymbol{v_i}^{\top})^{\top}$. The vector $\boldsymbol{\Lambda}_{i}(t_j)$ has a multivariate normal distribution with expectation $\boldsymbol{\mu_{\Lambda_{ij}}}$ and variance covariance matrix $\boldsymbol{V_{\Lambda_{ijj}}}=\text{var}\Big\{\boldsymbol{\Lambda}_{i}(t_j)\Big\}$, and the vector $\boldsymbol{\Lambda}_{i} = \left(\boldsymbol{\Lambda}_{i}(t_j)^\top\right)^\top_{j \in \tau}$ has a multivariate normal distribution with 														
expectation 														
$\boldsymbol{\mu_{\Lambda_i}} = \left(\boldsymbol{\mu_{\Lambda_{ij}}}^{\top}\right)^{\top}_{j \in \tau}$ and variance-covariance matrix $\boldsymbol{V_{\Lambda_i}} = \left(\boldsymbol{V_{\Lambda_{ijj'}}}\right)_{(j,j') \in \tau^2}$ where														
\begin{equation}\label{distribution_Lambda_1}														
\boldsymbol{\mu_{\Lambda_{ij}}}= E\Big\{\boldsymbol{\Lambda_i}(t_j)\Big\}= \boldsymbol{\Psi_{i, \delta}}(0,j,0)\boldsymbol{X^0_i} \boldsymbol{\beta} + \displaystyle \Big\{\delta \sum\limits_{s=1}^{j}\boldsymbol{\Psi_{i, \delta}}(0,j,s)\boldsymbol{X_i}(t_s)\boldsymbol{\gamma}\Big\}\mathds{1}_{\{j > 0\}} 						
\end{equation}  														
and 														
\begin{equation}\label{distribution_Lambda_2}														
\begin{array}{lll}														
\boldsymbol{V_{\Lambda_{ijj'}}} &=&\text{cov}\Big\{\boldsymbol{\Lambda_i}(t_j);\boldsymbol{\Lambda_i}(t_{ij'})\Big\}\\													
&=&\boldsymbol{\Psi_{i,\delta}}(0,j,0)\boldsymbol{\Psi_{i,\delta}}(0,j',0)^{\top} \\														
& & + \Big[\boldsymbol{\Psi_{i,\delta}}(0,j,0)\boldsymbol{B_{uv}}\Big\{\displaystyle \delta \sum\limits_{s'=1}^{j'}\boldsymbol{\Psi_{i,\delta}}(0,j',s')\boldsymbol{Z_i}(t_{is'})\Big\}^{\top}\Big] \mathds{1}_{\{j' > 0\}} \\											
& & + \Big[{\displaystyle\Big\{\delta\sum\limits_{s=1}^{j}\boldsymbol{\Psi_{i,\delta}}(0,j,s)\boldsymbol{Z_i}(t_s)\Big\} }\boldsymbol{B_{uv}}^{\top}\boldsymbol{\Psi}_{i,\delta}(0,j',0)^{\top}\Big]\mathds{1}_{\{j > 0\}} \\														
& & + \Big[ {\Big\{\displaystyle \delta\sum\limits_{s=1}^{j}\boldsymbol{\Psi}_{i,\delta}(0,j,s)\boldsymbol{Z}_{i}(t_s)\Big\}}\boldsymbol{B_{v}}\Big\{\displaystyle\delta\sum\limits_{s'=1}^{j'}\boldsymbol{\Psi}_{i,\delta}(0,j',s')\boldsymbol{Z}_{i}(t_{is'})\Big\}^{\top}\Big] \mathds{1}_{\{\min(j,j') > 0\}}										
\end{array}														
\end{equation}														
It can be deduced that the vector of incomplete and transformed data $\boldsymbol{\tilde{Y}^*_{ij}}$ at occasion $j$ is multivariate Gaussian with expectation $\boldsymbol{\mu_{\tilde{Y}^*_{ij}}}=\boldsymbol{M_{ij}P\mu_{\Lambda_{ij}}}$  and variance-covariance matrix \\ $\boldsymbol{V_{\tilde{Y}^*_{ij}}}= \boldsymbol{M_{ij}}\left(\boldsymbol{PV_{\Lambda_{ijj}}P}^{\top} + \boldsymbol{\Sigma}\right) \boldsymbol{M_{ij}}^{\top}$, and the total vector of incomplete and transformed data
$\boldsymbol{\tilde{Y}^{*}_{i}} = \left(\boldsymbol{\tilde{Y}^{*\top}_{ij} }\right)^{\top}_{j \in \tau_i}$ 
is multivariate Gaussian with expectation $\boldsymbol{\mu_{\tilde{Y}^*_i}} = {\left(\boldsymbol{\mu_{\tilde{Y}^{*}_{ij}}}^{\top}\right)^{\top}_{j \in \tau_i}}$ and variance-covariance matrix 
$\boldsymbol{V_{\tilde{Y}^*_i}}$, a block matrix with $\boldsymbol{M_{ij}P V_{\Lambda_{ijj'}} P}^{\top}\boldsymbol{M_{ij'}}^{\top} + \left(\boldsymbol{M_{ij} \Sigma M_{ij'}}^{\top} \right)\mathds{1}_{\{j=j'\}} $ the $(j,j')$ block.
\subsubsection{Likelihood}\label{Estimation_Distributions_Likelihood}
 As the $N$ subjects of the sample are independent, the log-likelihood of the model is $L(Y^*;\theta) = \displaystyle\sum_{i=1}^{N}\log \Big\{\mathcal{L}_i(\boldsymbol{Y_i^*}; \boldsymbol{\theta})\Big\}$ with  $\mathcal{L}_i(\boldsymbol{Y_i^*}; \boldsymbol{\theta})$ the individual contribution to the likelihood. Here,\\ $\boldsymbol{\theta} = \Big(\boldsymbol{\beta}^{\top}, \boldsymbol{\gamma}^{\top}, vec (\boldsymbol{L})^{\top}, (\boldsymbol{\alpha_{dd'}}^{\top})_{d,d' \in \{1, \dots, D\}^2}, (\sigma_k)_{k \in \{1,\dots, K\}}, \boldsymbol{\eta}^\top\Big)^\top$ is the whole vector of parameters. Using the Jacobian of the link functions $H_k$ ($k=1,...,K$), the individual contribution is:														
\begin{equation}\label{Likelihood_equ_1}														
\mathcal{L}_i(\boldsymbol{Y_i^*}; \boldsymbol{\theta}) = \phi_i(\boldsymbol{\tilde{Y}^*_i}; \boldsymbol{\mu_{\tilde{Y}^*_i}}, \boldsymbol{V_{\tilde{Y}^*_{i}}})\displaystyle\prod_{j\in \tau_i}\displaystyle\prod_{l=1}^{K_{ij}^*}\mathcal J_{\kappa(l)}\Big\{H_{\kappa(l)}(\widetilde{Y}^*_{ij\kappa(l)}; \boldsymbol{\eta_{\kappa(l)}})\Big\}														
\end{equation}														
where $\phi_{i}(.;\boldsymbol{\mu},\boldsymbol{V})$ denotes the density function of a multivariate Gaussian vector with expectation $\boldsymbol{\mu}$ and variance-covariance $\boldsymbol{V}$, and  $\mathcal J_{\kappa(l)}\Big\{H_{\kappa(l)}(\tilde{Y}^*_{ij\kappa(l)}; \boldsymbol{\eta_{\kappa(l)}})\Big\}$ denotes the Jacobian of the link function $H_{\kappa(l)}$ used to transform $Y^*_{ij\kappa(l)}$, the $l^{\text{th}}$ observed marker at occasion $j$ for subject $i$.
For instance, with the linear link function defined in \eqref{measurement_model_1}, $\mathcal J_{\kappa(l)} \Big\{H_{\kappa(l)}(\tilde{Y}^*_{ij\kappa(l)}; \eta_{\kappa(l)}) \Big\} = \frac{1}{\eta_{1\kappa(l)}}$.

\subsubsection{Optimization Algorithm and Implementation}\label{Estimation_Distributions_Optimisation}														
														
The maximum likelihood estimates are obtained using an extended Levenberg-Marquardt algorithm \citep{marquardt1963algorithm} because of its robustness and good convergence rate. 														
At each iteration $p$, if necessary, the Hessian matrix $\boldsymbol{H^{(p)}}$ is diagonal-inflated to obtain a positive definite matrix $\boldsymbol{H^{*(p)}}$ which is used to update the parameters $\boldsymbol{\theta^{(p+1)}} = \boldsymbol{\theta^{(p)}} - \nu {\Big(\boldsymbol{H^{*(p)}}\Big)}^{-1}\boldsymbol{U}(\boldsymbol{\theta^{(p)}})$, with $\boldsymbol{U}(\boldsymbol{\theta^{(p)}})$ the gradient at iteration $p$ and $\nu$ the improvement control parameter. Convergence is reached 														
when $||\boldsymbol{\theta^{(p+1)}} - \boldsymbol{\theta^{(p)}}||_{2} < \epsilon_\theta$, $|L(\boldsymbol{Y^*}; \boldsymbol{\theta^{(p+1)}}) - L(\boldsymbol{Y^*}; \boldsymbol{\theta^{(p)}})| <\epsilon_L$ and $\frac{\boldsymbol{U}(\boldsymbol{\theta^{(p)}})^{\top}{(\boldsymbol{H^{(p)}})}^{-1}\boldsymbol{U}(\boldsymbol{\theta^{(p)})}}{n_{\text{para}}} < \epsilon_H$, with $n_{\text{para}}$ the total number of parameters. The latter criterion is by far the most stringent one and specifically targets maximum search so that $\epsilon_\theta=\epsilon_L=\epsilon_H=10^{-3}$ is small enough to ensure convergence. The variances of the  estimators are obtained from the inverse of $\boldsymbol{H^{(p)}}$. 														
														
Given the possibly high number of parameters, we first estimate the parameters for each process taken separately, then we start the maximization of the likelihood of the multivariate model from these simple estimates, setting initial values of the inter-dimension parameters to zero. 
The model estimation is implemented in R; it combines R and $C_{++}$ languages, and includes parallel computations.	The program is available for download at \url{https://github.com/bachirtadde/CInLPN}.											
														
\subsection{Marginal and subject-specific predictions}\label{Goodness-of-Fit}														
 The goodness-of-fit of the model can be assessed by comparing predictions with observations of the markers in their transformed scales. From notations defined in Section \ref{Estimation_Distributions}, marginal and conditional distributions of the markers are:														
\begin{equation}														
\boldsymbol{\tilde{Y}_i} \sim \mathcal{N}\Big( \boldsymbol{P\mu_{\Lambda_i}} ~, ~ \left( \boldsymbol{PV_{\Lambda_{i}} P}^{\top} + \boldsymbol{\Sigma_i}  \right) \Big).														
\end{equation} 														
\begin{equation}														
\left. \boldsymbol{\tilde{Y}_i}\right\rvert_{\boldsymbol{\Lambda_i}} \sim \mathcal{N}\left( \boldsymbol{P\Lambda_i}, ~ \boldsymbol{\Sigma_i} \right)														
\end{equation}														
														
where $\boldsymbol{\Sigma_i} $  is the block-diagonal matrix constituted of $n_i$ $\boldsymbol{\Sigma}$ blocks. 														
														
The marginal ($\boldsymbol{Y_i^{(M)}}$) and subject-specific ($\boldsymbol{Y_i^{(SS)}}$) predictions in the transformed scales are then respectively obtained by taking the expectations of the marginal and conditional distributions of the transformed markers at the parameter estimates $\hat{\theta}$ and at the predicted latent processes $\boldsymbol{\hat{\Lambda}_i}$ of $\boldsymbol{\Lambda_i}$ given the observations $\boldsymbol{\tilde{Y}_i^*}$:														
$														
\boldsymbol{\hat{\Lambda}_i} = E\left( \left. \boldsymbol{\Lambda_i}\right\rvert { \boldsymbol{\tilde{Y}_i^*}} \right) = \boldsymbol{\mu_{\Lambda_i}} + \boldsymbol{C_{\Lambda_i \tilde{Y}_i^*}} \boldsymbol{V^{-1}_{\tilde{Y}_i^*}}\left(\boldsymbol{\tilde{Y}_i^*} - \boldsymbol{\mu_{\tilde{Y}_i^*}} \right),														
$														
where $\boldsymbol{C_{\Lambda_i \tilde{Y}_i^*}} = \left( \boldsymbol{V_{\Lambda_{ijj'}} P}^{\top}\boldsymbol{M_{ij'}}^{\top} \right)_{(j,j') \in \tau^2}$ is the covariance matrix between $\boldsymbol{\Lambda_i}$ and $\boldsymbol{\tilde{Y}_i^*}$.														
														
Using these individual predictions, one can graphically compare either the marginal predictions $\boldsymbol{Y^{(M)}}$ or subject-specific predictions $\boldsymbol{Y^{(SS)}}$ averaged within time intervals to the observations averaged within the same time intervals. Marginal and subject-specific predictions in the natural scale of the markers can also be derived from the marginal and conditional distributions by using a Monte-Carlo approximation \citep{proust2013analysis}. 														
														
%----------------------------------------------														
\section{Simulations}\label{Simulations}									
We performed two series of simulations to evaluate the estimation program and the impact of time discretization on the interpretation of $\boldsymbol{A_{i,\delta}}(t)$ matrix.														
\subsection{Simulation Study 1: Validation of the estimation program}													
\subsubsection{Design}\label{Simulations_Design}														
To evaluate the estimation program, we generated a system of two Gaussian processes ($(\boldsymbol{\Lambda^1}(t_j))_{t_j \geq 0}$ and $(\boldsymbol{\Lambda^2}(t_j))_{t_j \geq 0}$), with $ t_j = j \times \delta $, $j \in \{0, \dots, J \}$. Each process is  measured by one longitudinal marker ($\boldsymbol{Y_1}$ and $\boldsymbol{Y_2}$, respectively) and two covariates, one continuous $C_1$ and one binary $C_2$. 														
We considered two scenarios: a covariate-specific structure of temporal influences (Scenario 1) and a time-dependent temporal influences structure (Scenario 2). In Scenario 1, we assumed a constant rate of change for the system of latent processes (with random intercepts and simple effects of both covariates in the sub-models for the initial level and the change over time) and a structure of temporal influences $\boldsymbol{A_{i,\delta}}$ different for each level of $C_1$:					
\begin{equation}\label{model_simu_1}									
\begin{array}{rl}														
 & \left\{														
\begin{array}{rl}														
 \Lambda_{i}^{1}(0) & = \beta_0^1 + \beta_1^1 C_{1,i} + \beta_2^1 C_{2,i} + u_i^1 \\														
\Lambda_{i}^{2}(0) & = \beta_0^2 + \beta_1^2 C_{1,i} + \beta_2^2 C_{2,i} + u_i^2 \\														
\frac{\Delta\Lambda^{1}_{i}(t_j +\delta)}{\delta} & = \gamma_0^1 + \gamma_1^1 C_{1,i} + \gamma_2^1 C_{2,i} + v_i^1 + (\alpha_{i,11}^{0} + \alpha_{i,11}^{1}C_{2,i}) \Lambda_{i}^{1}(t_j) +  (\alpha_{i,12}^{0} + \alpha_{i,12}^{1}C_{2,i})\Lambda_{i}^{2}(t_j)\\						
\frac{\Delta\Lambda^{2}_{i}(t_j +\delta)}{\delta} & = \gamma_0^2 + \gamma_1^2 C_{1,i} + \gamma_2^2 C_{2,i} + v_i^2 + (\alpha_{i,21}^{0} + \alpha_{i,21}^{1}C_{2,i})\Lambda_{i}^{1}(t_j) +  (\alpha_{i,22}^{0} + \alpha_{i,22}^{1}C_{2,i})\Lambda_{i}^{2}(t_j)							
\end{array}														
\right.														
\\ & \frac{Y_{ijk}-\eta_{0k}}{\eta_{1k}} = \Lambda_{i}^k(t_j) + \tilde{\epsilon}_{ijk}, ~~~~ k = 1,2			
\end{array}														
\end{equation}														
where $\boldsymbol{u_i}= (u_i^1, u_i^2)^{\top}$, $\boldsymbol{v_i}= (v_i^1, v_i^2)^{\top}$ and  $(\boldsymbol{u_i}^{\top},\boldsymbol{v_i}^{\top})^{\top} \sim \mathcal{N}( 0, \boldsymbol{LL}^{\top})$ with $\boldsymbol{L}$ such that the random effects are independent between dimensions, and $\tilde{\epsilon}_{ijk} \sim \mathcal{N}( 0, \sigma_k^2$), $\forall k  \in \{1,2\} $.% $\forall k,k' \in \{1,2\}^2$. 														
														
In scenario 2, we considered initial levels adjusted for the binary covariate $C_2$, constant rates of change with no adjustment, and temporal influences between two processes $\boldsymbol{A_{i,\delta}}(t_j)$ that evolved with time: 
\begin{equation}\label{model_simu_2}								
\begin{array}{rl}												
& \left\{														
\begin{array}{rl}													
 \Lambda_{i}^{1}(0) & = \beta_0^1 + \beta_1^1 C_{2,i} + u_i^1 \\
\Lambda_{i}^{2}(0) & = \beta_0^2 + \beta_1^2 C_{2,i} + u_i^2 \\		
\frac{\Delta\Lambda^{1}_{i}(t_j + \delta)}{\delta} & = \gamma_0^1  + v_i^1 + a_{11}(t_j)\Lambda_{i}^{1}(t_j) +  a_{12}(t_j)\Lambda_{i}^{2}(t_j)\\	\frac{\Delta\Lambda^{2}_{i}(t_j + \delta)}{\delta} & = \gamma_0^2 + v_i^2 + a_{21}(t_j)\Lambda_{i}^{1}(t_j) +  a_{22}(t_j)\Lambda_{i}^{2}(t_j)\\	\end{array}														
\right.														
\\  & \frac{Y_{ijk}-\eta_{0k}}{\eta_{1k}} = \Lambda_{i}^k(t_j) + \tilde{\epsilon}_{ijk}, ~~~~ k = 1,2,								
\end{array}														
\end{equation}													
where $\boldsymbol{u_i}= (u_i^1, u_i^2)^{\top}$, $\boldsymbol{v_i}= (v_i^1, v_i^2)^{\top}$ and  $(\boldsymbol{u_i}^{\top},\boldsymbol{v_i}^{\top})^{\top} \sim 														
\mathcal{N}( 0, \boldsymbol{LL}^\top)$ with $\boldsymbol{L}$ such that the random effects are independent between dimensions, and $\tilde{\epsilon}_{ijk} \sim \mathcal{N}( 0, \sigma_k^2$), $\forall~k  \in \{1,2\}$. Each element of the matrix of temporal influences $a_{kk'}(t)$ is defined from a basis of quadratic B-splines with one internal knot at the median $(\boldsymbol{S_m})_{\{m=1,3\}}$ so that $a_{kk'}(t) = \alpha_{kk'}^0 + \alpha_{kk'}^{1}\boldsymbol{S_1}(t) + \alpha_{kk'}^{2}\boldsymbol{S_2}(t) + \alpha_{kk'}^{3}\boldsymbol{S_3}(t) $, $\forall ~k\neq k'~\text{and}~(k,k') \in \{1,2\}^2$ (diagonal elements were not adjusted). 														
														
The design of the simulations and the parameters were chosen to mimic the ADNI data. Dimensions 1 and 2 were cerebral anatomy and cognitive ability, each one measured by a specific composite score. Covariate $C_1$ represented the baseline age centered on the mean age in decade and was generated according to a Gaussian distribution: $C_1 \sim \mathcal{N}(0,~0.64)$. Covariate $C_2$ corresponded to the indicator of group CN \emph{versus} group MCI. It was generated according to a Bernoulli distribution with probability 0.37. 														
We used a discretization step $\delta = 1$ (i.e., 6 months in ADNI 1 data). Markers observations were generated every 6 months up to 3 years. Thus a subject had 7 repeated measures at occasions $j \in \{ 0, 1, \dots, 6\}$. We also considered a design in which scheduled visits could be missed completely at random with a probability of 0.15, and when a visit was not missed, a marker could be missing with a probability of 0.07.												
For each design and scenario, we generated 1000 samples of 512 subjects.								%					
\subsubsection{Results}														
Web Table 1 of the Web Appendix B and Table \ref{table1} provide the results of the simulations for scenarios 1 and 2, respectively. In both settings, all the parameters were correctly estimated with satisfying coverage rates in the absence of missing data (left part) and in the presence of missing data (right part). %\cpl{nécessaire? dit dans le paragraphe précédent: In the latter, 15\% of occasions were missing and 7\% of outcomes were missing for non-missing occasions.} 								
\begin{table}[p]
\small
\caption{Results of the simulations for scenario 2 (1000 replicates of samples of size 512).} \label{table1}
\centering
  \begin{threeparttable}											
\centering														
\begin{tabular}{rrrrrrrrrrrrr}														
\hline														
& & \multicolumn{5}{c}{without missing values} & & \multicolumn{5}{c}{with missing values$^\star$}\\
  \cline{3-7}  \cline{9-13} 														
& $\theta$ & $\hat{\theta}$ & bias$^{\dagger}$ & ESE$^{\ddagger}$ & ASE$^{\ddagger}$ & CR(\%) & & $\hat{\theta}$ & bias & ESE$^{\ddagger}$ & ASE$^{\ddagger}$ & CR(\%)\\ 
 \cline{1-2} \cline{3-7} \cline{9-13}														

  $\beta_1^1$     & -1.635 & -1.644 & 0.5  & 0.109 & 0.108 & 95.5 && -1.644 & 0.5  & 0.109 & 0.109 & 94.7 \\ 
  $\beta_1^2$     & -1.784 & -1.789 & 0.2  & 0.115 & 0.119 & 95.7 && -1.790 & 0.3  & 0.119 & 0.121 & 94.8 \\ 
  $\gamma_0^1$    & 0.009  & 0.009  & 1.4  & 0.009 & 0.009 & 96.0 && 0.009  & 1.6  & 0.009 & 0.010 & 95.5 \\ 
  $\gamma_0^2$    & -0.053 & -0.053 & 1.5  & 0.017 & 0.017 & 95.5 && -0.053 & 1.3  & 0.017 & 0.018 & 95.6 \\ 
  L(3,1)          & 0.032  & 0.031  & 3.2  & 0.009 & 0.010 & 95.2 && 0.031  & 3.2  & 0.010 & 0.010 & 95.5 \\ 
  L(4,2)          & -0.011 & -0.009 & 19.5 & 0.014 & 0.015 & 95.4 && -0.009 & 20.5 & 0.014 & 0.015 & 94.7 \\ 
  L(3,3)          & 0.094  & 0.094  & 0.1  & 0.006 & 0.006 & 94.6 && 0.094  & 0.3  & 0.006 & 0.006 & 93.9 \\ 
  L(4,4)          & 0.169  & 0.169  & 0.2  & 0.010 & 0.011 & 95.0 && 0.169  & 0.1  & 0.011 & 0.011 & 94.7 \\ 
  $\alpha_{11}^0$ & -0.012 & -0.011 & 8.2  & 0.008 & 0.009 & 95.3 && -0.011 & 8.3  & 0.009 & 0.009 & 95.6 \\ 
  $\alpha_{12}^0$ & 0.115  & 0.114  & 0.9  & 0.018 & 0.018 & 93.8 && 0.113  & 1.1  & 0.020 & 0.020 & 94.9 \\ 
  $\alpha_{12}^1$ & -0.092 & -0.092 & 0.3  & 0.027 & 0.027 & 94.9 && -0.092 & 0.6  & 0.030 & 0.030 & 96.4 \\ 
  $\alpha_{12}^2$ & -0.028 & -0.028 & 3.2  & 0.015 & 0.015 & 94.5 && -0.027 & 4.1  & 0.017 & 0.017 & 95.1 \\ 
  $\alpha_{12}^3$ & -0.069 & -0.069 & 0.1  & 0.026 & 0.026 & 94.1 && -0.069 & 0.2  & 0.028 & 0.029 & 95.1 \\ 
  $\alpha_{21}^0$ & 0.134  & 0.135  & 0.7  & 0.033 & 0.034 & 94.2 && 0.135  & 0.4  & 0.036 & 0.036 & 94.8 \\ 
  $\alpha_{21}^1$ & -0.076 & -0.075 & 0.9  & 0.052 & 0.052 & 93.6 && -0.075 & 1.2  & 0.058 & 0.058 & 94.5 \\ 
  $\alpha_{21}^2$ & 0.024  & 0.022  & 11.3 & 0.031 & 0.035 & 95.4 && 0.023  & 7.9  & 0.034 & 0.034 & 95.4 \\ 
  $\alpha_{21}^3$ & -0.140 & -0.130 & 6.8  & 0.052 & 0.052 & 94.2 && -0.130 & 7.0  & 0.058 & 0.058 & 94.0 \\ 
  $\alpha_{22}^0$ & 0.009  & 0.007  & 22.2 & 0.012 & 0.012 & 94.8 && 0.007  & 21.6 & 0.013 & 0.013 & 94.7 \\ 
  $\sigma_1$      & 0.376  & 0.378  & 0.5  & 0.013 & 0.013 & 94.9 && 0.378  & 0.5  & 0.013 & 0.013 & 95.2 \\ 
  $\sigma_2$      & 0.686  & 0.688  & 0.2  & 0.027 & 0.026 & 93.9 && 0.688  & 0.3  & 0.028 & 0.027 & 93.6 \\ 
  $\eta_{01}$     & 3.878  & 3.886  & 0.2  & 0.196 & 0.198 & 95.1 && 3.885  & 0.2  & 0.196 & 0.199 & 94.8 \\ 
  $\eta_{11}$     & 2.678  & 2.667  & 0.4  & 0.087 & 0.087 & 95.1 && 2.667  & 0.4  & 0.088 & 0.088 & 94.9 \\ 
  $\eta_{02}$     & 2.589  & 2.591  & 0.1  & 0.112 & 0.114 & 95.9 && 2.591  & 0.1  & 0.114 & 0.115 & 95.3 \\ 
  $\eta_{12}$     & 1.472  & 1.470  & 0.1  & 0.054 & 0.052 & 93.5 && 1.470  & 0.2  & 0.056 & 0.054 & 93.4 \\ 
   \hline														
\end{tabular}	
\begin{tablenotes}
\item $^\star$ (15\% missing occasions, 7\% missing outcomes),	\item $^{\dagger}$ relative bias(\%),					
\item $^{\ddagger}$ ASE is the asymptotic standard error, ESE is the empirical standard error and CR is the coverage rate of the 95\% confidence interval,
\item diagonal elements of the temporal influences matrix were not adjusted for time.
\end{tablenotes}
\end{threeparttable}
\end{table}																						
														
\subsection{Simulation Study 2: Impact of the discretization step on the temporal influence structure between processes}														
\subsubsection{Design}														
To formally assess whether the interpretation under the discretized time was the same as the one obtained under continuous time, we assessed the type-I error rate associated with each non diagonal element of the matrix of temporal influences $\boldsymbol{A}$ under three discretization steps $\delta = 1/3$ (for 2 months), $1/2$ (for 3 months) and $1$ (for 6 months) when data were actually generated under continuous time (approximated by a step $\delta$=0.001). 														
We considered for this a system of three latent processes ($(\boldsymbol{\Lambda^1}(t_j))_{t_j \geq 0}$, $(\boldsymbol{\Lambda^2}(t_j))_{t_j \geq 0}$, and $(\boldsymbol{\Lambda^3}(t_j))_{t_j \geq 0}$ ), with $t_j = j \times \delta, ~ \delta = 0.001$. Each process is measured by one Gaussian repeated marker ($\boldsymbol{Y_1}$, $\boldsymbol{Y_2}$, $\boldsymbol{Y_3}$). The processes have constant rate of change, with no adjustment for covariates and a matrix of temporal influences $\boldsymbol{A}$ constant over time: 				
\begin{equation}\label{model_simu_3}						
\begin{array}{rl}											
& \left\{														
\begin{array}{rl}													
\Lambda_{i}^{1}(0) & = \beta_0^1  + u_i^1 \\			
\Lambda_{i}^{2}(0) & = \beta_0^2 + u_i^2 \\					
\Lambda_{i}^{3}(0) & = \beta_0^3 + u_i^3 \\					
\frac{\Delta\Lambda^{1}_{i}(t_j + \delta)}{\delta} & = \gamma_0^1 + v_i^1 + a_{11}\Lambda_{i}^{1}(t_j) +  a_{12}\Lambda_{i}^{2}(t_j) + a_{13}\Lambda_{i}^{3}(t_j)\\		
\frac{\Delta\Lambda^{2}_{i}(t_j + \delta)}{\delta} & = \gamma_0^2 + v_i^2 + a_{21}\Lambda_{i}^{1}(t_j) +  a_{22}\Lambda_{i}^{2}(t_j) + a_{23}\Lambda_{i}^{3}(t_j)\\		
\frac{\Delta\Lambda^{3}_{i}(t_j + \delta)}{\delta} & = \gamma_0^3 + v_i^3 + a_{31}\Lambda_{i}^{1}(t_j) +  a_{32}\Lambda_{i}^{2}(t_j) + a_{33}\Lambda_{i}^{3}(t_j)\\		
\end{array}													
\right.\\													
& \frac{Y_{ijk}-\eta_{0k}}{\eta_{1k}} = \Lambda_{i}^k (t_j) + \tilde{\epsilon}_{ijk}, ~~~~~ k = 1,2,3						
\end{array}												
\end{equation}

The simulation model mimicked the ADNI data with cerebral anatomy, cognitive ability and functional autonomy as dimensions, respectively measured by a specific composite score. We estimated the model on the ADNI data with $\delta$=1 (6 months) and transformed the parameters to the scale $\delta$=0.001 to generate the data in almost continuous time. We provide in the Web Appendix A the formulas to relate model components defined under $\delta$=1 and $\delta$=0.001.														
Elements of the matrix of temporal influences were set one by one to 0 in $\delta = 0.001$ scale to evaluate the associated type-I error rate. Latent processes were generated with solver from dsolve package \citep{Soetaert2010desolve} and observations were derived every 6 months up to 3 years. We considered each time 1000 samples of 512 subjects.

\subsubsection{Results}\label{Simulations_Results}									
Table \ref{table2} displays the type-I error rates (in percentage) associated with each non diagonal element of the matrix of temporal influences $\boldsymbol{A}$ when estimated with discretization steps $\delta$=1/3, $1/2$ and $1$. All the type-I error rates are close to the nominal 5\% rate (95\% interval of expected rates of $[3.6, ~ 6.4]$ with 1000 replicates). We note however that with a discretization step of $\delta$=1, the type-I error rates begin to somewhat increase, suggesting that with larger discretization steps, causal interpretations are altered.  

\begin{table}[p]														
\footnotesize														
\caption{Type-I error rates (in \%) associated with each non diagonal element of the matrix of temporal influences $\boldsymbol{A}$ when the matrix of temporal influences is generated approximately in  continuous time ($\delta$=0.001) and estimated with discretization steps: $\delta$=1/3, $\delta$=1/2, $\delta$=1 for 2, 3 and 6 months respectively (1000 replicates; expected 95\% interval $[3.6, ~ 6.4]$ for the nominal type-I error of 5\%). \vspace{2mm}} \label{table2}							
\centering														
\centering									
\begin{tabular}{lccc}														
 \hline														
 &$\delta=1/3$ & $\delta=1/2$ &  $\delta=1$ \\ \cline{2-4}					
 Parameter & \multicolumn{3}{c}{$\hat{r}$}\\			
 \hline														
$a_{12}$ & 5.4 & 4.9 & 6.5   \\
$a_{13}$ & 5.3 & 5.3 & 8.6   \\
$a_{21}$ & 5.6 & 6.0 & 7.5   \\
$a_{23}$ & 4.7 & 6.0 & 4.1  		    \\
$a_{31}$ & 5.4 & 5.2 & 7.1   \\
$a_{32}$ & 4.5 & 4.8 & 4.8  		    \\												
\hline														
\end{tabular}				

\end{table}	

\section{Application}\label{Application}

\begin{figure}														
\centering														
\includegraphics[width=1.1\textwidth]{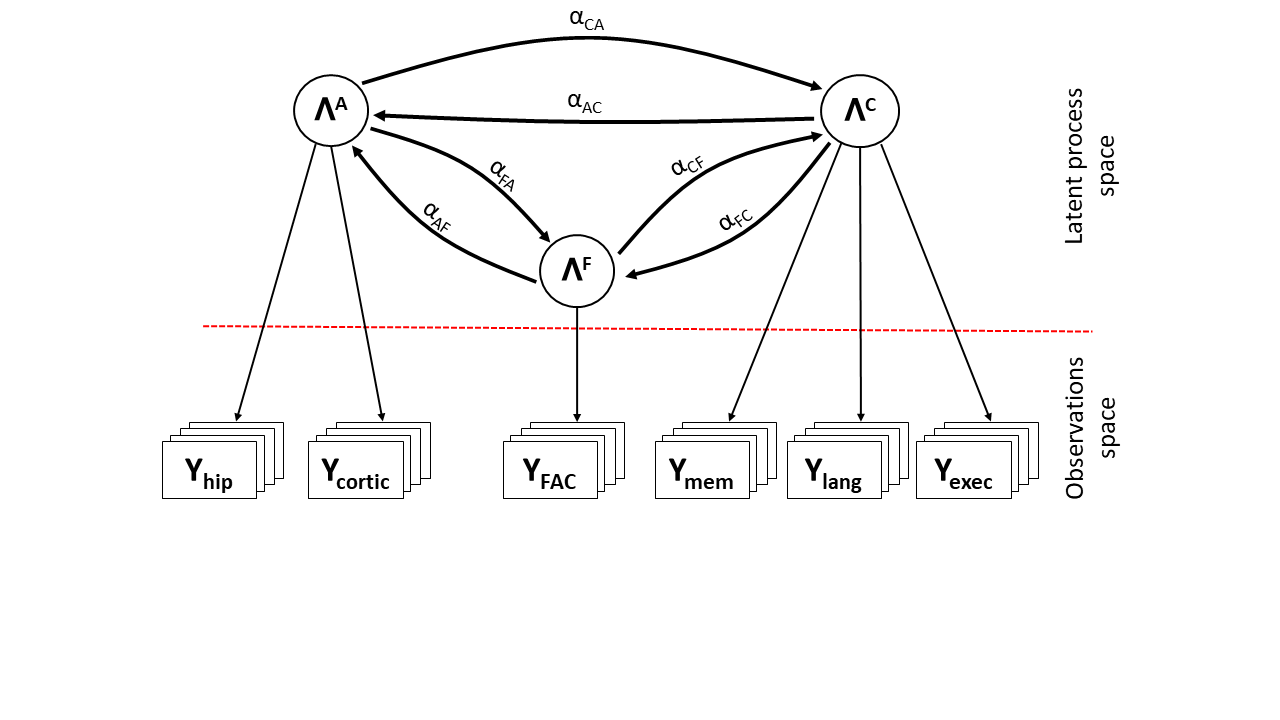}														
\caption{Graph of the dynamic causal model considered on ADNI 1 data with three dimensions (labelled $\boldsymbol{\Lambda^A}$ for cerebral anatomy, $\boldsymbol{\Lambda^C}$ for cognitive ability and $\boldsymbol{\Lambda^F}$ for functional autonomy). Cerebral anatomy is measured by two repeated volumes ($Y_{\text{hip}}$ and $Y_{\text{cortic}}$ for hippocampal volume and mean regional cortical thickness, respectively), cognitive ability by three repeated scores ($Y_{\text{mem}}$, $Y_{\text{lang}}$ and $Y_{\text{exec}}$ for sumscores in memory, language and executive functions, respectively) and functional autonomy by one repeated score ($Y_{\text{FAQ}}$ for FAQ scale). }														
\label{appl_figure1}														
\end{figure}

Using the ADNI data (Section \ref{data}), the application aimed to describe the decline over time of cerebral anatomy, cognitive ability and functional autonomy in three clinical stages of AD (normal aging (CN), Mild Cognitive Impairment (MCI) and diagnosed with Alzheimer's Disease (dAD)) and to quantify the temporal influences between these dimensions by assessing especially whether the relationships differ according to the clinical stage. The dynamic model applied on ADNI is summarized in Figure \ref{appl_figure1}.

%---------------------------------------------------------									 
\subsection{Measures and ADNI sample description}
Cerebral anatomy ($\boldsymbol{\Lambda^A}$) was defined as the process underlying the hippocampal volume relative to total intracranial volume and the mean cortical thickness of nine regions (Freesurfer version 4.4.0 for longitudinal data) previously identified in a cortical signature of AD \citep{dickerson2008cortical}. Global cognitive ability ($\boldsymbol{\Lambda^C}$) was defined as the process underlying three cognitive markers of memory, language and executive functioning, respectively. Each marker was a composite score previously identified from the psychometric tests available in ADNI \citep{park2012confirmatory}. Functional autonomy ($\boldsymbol{\Lambda^F}$) was defined as the process underlying the observed sumscore of the FAQ (Functional Assessment Questionnaire) composed of 30 items \citep{pfeffer_measurement_1982}.
Covariates were the age at entry (centered around 75.4 and indicated in decades), gender, educational level (low level if $\leq$ 12 years \emph{vs} high level if $>$12 years), the APOE genotype ($\epsilon$4 carrier \emph{vs} $\epsilon$4 non-carrier ) and the 3 clinical stages (CN, MCI, dAD) as defined at inclusion in ADNI (we did not consider possible changes in stages during follow-up). We selected in the sample all the subjects who had no missing data for the covariates and had at least one observation for each dimension during the follow-up. The main analysis included 656 subjects (82\% of the initial sample). The sample consisted of 190 (29\%) subjects at healthy stage (CN), 322 (49\%) subjects at mild cognitive impairment stage (MCI), and 144 (22\%) subjects diagnosed with Alzheimer's disease (dAD); 43\% were females, 83\% had a high educational level and 49\% carried APOE $\epsilon $4 allele. The mean age at entry was 75.4 years old (sd=6.6). The mean number of visits was 5, 6 and 4 for CN, MCI and dAD subjects, respectively.						
%---------------------------------------------------------													
\subsection{Specification of the dynamic multivariate model}														
											
%The dynamic model applied to ADNI 1 is summarized in Figure \ref{appl_figure1}. The cerebral anatomy ($\boldsymbol{\Lambda^A}$) and cognitive ability ($\boldsymbol{\Lambda^C}$) dimensions were repeatedly measured by composite scores $\boldsymbol{Y_A}$ and $\boldsymbol{Y_C}$ respectively. The functional autonomy dimension was repeatedly measured by the standardized sum-score of 30 items, $\boldsymbol{Y_F}$.
The initial levels of the processes were adjusted for gender, education, APOE genotype, clinical stage, and age at baseline. Changes of the processes over time were adjusted for gender, education, APOE genotype, clinical stage. In order to account for the correlations between individual repeated measures, we included random intercepts on the initial levels of the processes (correlated between processes) and random intercepts on the changes of the processes over time (independent between processes). 
The matrix of temporal influences that captured the interrelations between processes was constant over time and adjusted for clinical stage: $a_{dd'} = \alpha^0_{dd'} + MCI \times \alpha^1_{dd'} + dAD \times \alpha^2_{dd'}$, for $(d,d') \in \{A,C,F\}^2$.													
To correct the possible departure from normality of each observed marker, we transformed them using integrated quadratic splines (as in equation \eqref{transformation}) with 2 internal knots at the terciles for scores measuring cerebral and cognitive dimensions and with one internal knot at the median for the functional score.

In the main analysis, we considered a discretization step of approximately 3 months ($\delta = 0.23$ year) although observations were sparser as scheduled every 6 months. The modelling strategy consisted in finding first the best adjustment for each process taken separately with a significance threshold for covariate effects at 25\%. Then, the whole multivariate model was estimated. In secondary analyses, we re-estimated the final model by considering a discretization step of 1.5 months ($\delta = 0.125$ year) in order to evaluate whether the interpretations varied with a smaller step. 														
													
%---------------------------------------------------------												
\subsection{Results}\label{appli:result}														
														
\subsubsection{Latent process-specific trajectories}														
														
Estimates of fixed effects, Cholesky's decomposition parameters (for the random effects variance-covariance matrix) and measurement model parameters are provided in Web Tables 2, 3 and 4 of the Web Appendix C. In summary, older age, male gender, APOE $\epsilon 4$ carrying, and clinical stages MCI and dAD were associated with lower cerebral anatomy and lower cognitive ability levels at baseline. Higher education was associated with lower cerebral anatomy level but associated with higher cognitive ability level at baseline. Only clinical stages MCI and dAD were associated with lower functional autonomy level at baseline. 														
Adjusted for other dimension levels, APOE $\epsilon4$ carrying was associated with steeper declines in cerebral anatomy, cognitive ability and functional autonomy;
%after adjustment for other dimensions. Clinical stage dAD was associated with steeper cerebral anatomy decline and both clinical stages MCI and dAD were associated with steeper declines in cognitive ability and functional autonomy. 
and higher education was associated with steeper cerebral anatomy decline and smaller cognitive ability decline. Web Figure 1 of the Web Appendix C depicts the expected trajectories of each dimension according to stage for two profiles of individuals (women noncarrier of APOE $\epsilon$4 and with lower educational level; men carrier of APOE $\epsilon4$ and with higher educational level).

\subsubsection{Temporal influence structure between processes}									
\begin{figure}[p]														
\centering														
\includegraphics[width=1.3\textwidth]{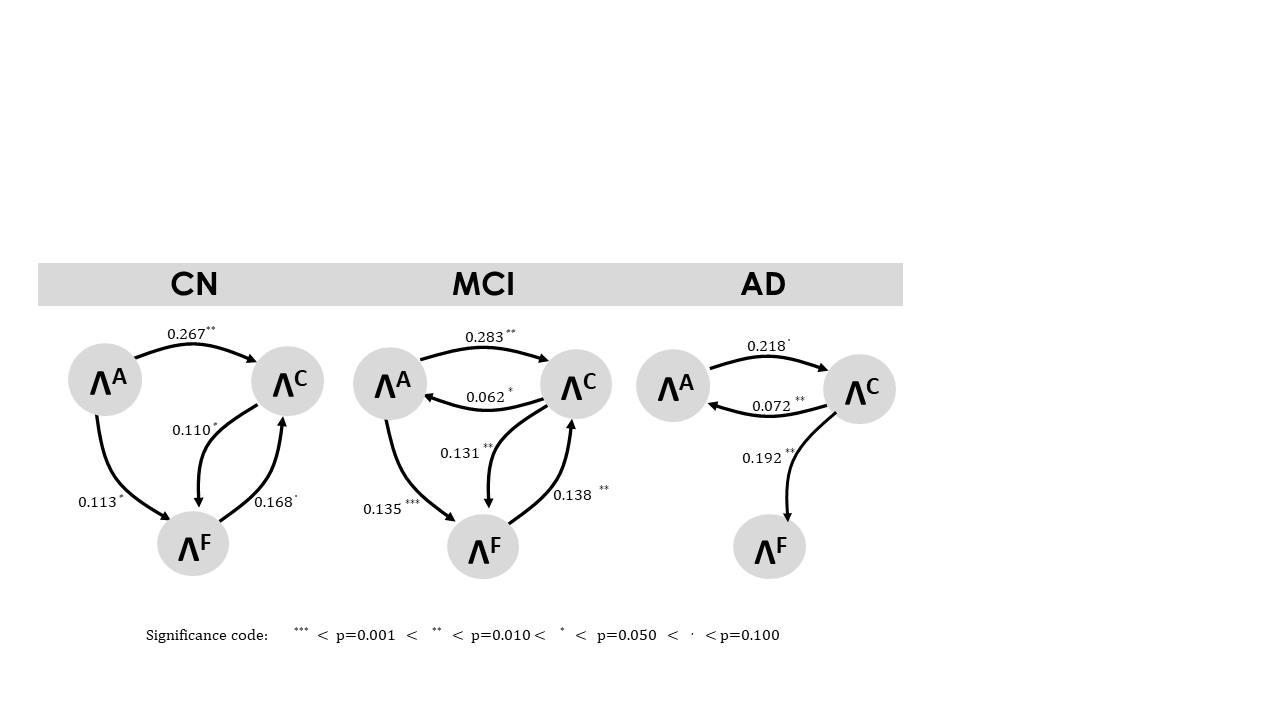}													
\caption{Temporal relationships estimated between cerebral anatomy ($\boldsymbol{\Lambda^A}$), cognitive ability ($\boldsymbol{\Lambda^C}$), and functional autonomy ($\boldsymbol{\Lambda^F}$) at healthy (CN), MCI and dAD stages. Arrows represent effects of one dimension on the change of another dimension. Only the effects identified from Table \ref{appl_estimates_CS} with a significance level lower than 10\% are reported: numbers indicate the estimate and stars indicate the level of significance according to the Wald test.}	
\label{appl_figure2}
\end{figure}

Estimates of the matrix of temporal influences are given in Table \ref{appl_estimates_CS} and summarized in Figure \ref{appl_figure2} according to stage. Figure \ref{appl_figure2} shows that the temporal influences between cerebral anatomy ($\boldsymbol{\Lambda^A}$), cognitive ability ($\boldsymbol{\Lambda^C}$), and functional autonomy ($\boldsymbol{\Lambda^F}$) evolve from healthy stage to Alzheimer's disease stage. At normal stage, cerebral dimension affects significantly the change of cognitive and functional dimensions {while cognitive and functional dimensions tend to have reciprocal temporal dependencies}. From MCI stage, the three dimensions become strongly interdependent with cognition as the central point: reciprocal temporal relationships appear between the anatomical and cognitive dimensions in addition to those between the functional and cognitive dimensions. At the dAD stage, the influence of the cerebral anatomy on functional autonomy and the effect of functional dimension (probably through its social component) on the cognitive dimension are attenuated and no more significant.    \\														
When considering a discretization step 1.5 months instead of 3 months, the Akaike Information criterion did not substantially differ (AIC = 29550.5 with 3 months and AIC = 29547.1 with 1.5 months) and the results regarding the temporal influence relationships remained the same (see Web Table 5 of the Web Appendix C).

%-----------------------------------------														
														
%---------------------------------------------														
\begin{table}[p]														
\small														
\centering														
\caption{Estimates of the temporal influences between cerebral anatomy, cognitive ability and functional autonomy in ADNI1 using a discretization step of 3 months. %Effects are adjusted for clinical stage (CN, MCI, dAD). 
\vspace{1mm} }														
\label{appl_estimates_CS}														
\begin{tabular}{lrrrrr}														
\hline														
 & \multicolumn{2}{c}{Parameter} & Estimate & SE & p-value \\ 														
\hline
\multicolumn{3}{l}{\bf{Influence on cerebral anatomy:}} & & &\\							
Effect of cerebral anatomy 	&	Intercept	&	$\alpha^0_{AA}$	&	-0.145	&	0.034	&	$<0.001$	\\	
                             	& MCI 		&	$\alpha^1_{AA}$	&	0.020	&	0.018	&	0.271	\\	
                            	& dAD 		&	$\alpha^2_{AA}$	&	0.002	&	0.025	&	0.946	\\	\cline{2-6}
                            	&	Intercept	&	$\alpha^0_{AC}$	&	0.014	&	0.015	&	0.377	\\	
Effect of cognitive ability 	& MCI 		&	$\alpha^1_{AC}$	&	0.048	&	0.018	&	0.008	\\	
                            	& dAD 		&	$\alpha^2_{AC}$	&	0.058	&	0.027	&	0.030	\\	\cline{2-6}
	                                                          											
                            	&	Intercept	&	$\alpha^0_{AF}$	&	0.015	&	0.031	&	0.627	\\	
Effect of functional  autonomy 	& MCI 	&	$\alpha^1_{AF}$		&	-0.023	&	0.032	&	0.478	\\	
                                	& dAD 		&	$\alpha^2_{AF}$	&	-0.040	&	0.037	&	0.286	\\	\hline
												
\multicolumn{3}{l}{\bf{Influence on cognitive ability:}} & & &\\												
												
                            	&	Intercept	&	$\alpha^0_{CA}$	&	0.267	&	0.078	&	$<0.001$	\\	
Effect of cerebral anatomy 	& MCI 		&	$\alpha^1_{CA}$	&	0.016	&	0.060	&	0.792	\\	
                            	& dAD 		&	$\alpha^2_{CA}$	&	-0.049	&	0.080	&	0.535	\\	\cline{2-6}
                            	&	Intercept	&	$\alpha^0_{CC}$	&	-0.548	&	0.148	&	$<0.001$	\\	
Effect of cognitive ability	 & MCI 		&	$\alpha^1_{CC}$	&	0.171	&	0.055	&	0.002	\\	
                            	& dAD 		&	$\alpha^2_{CC}$	&	0.115	&	0.076	&	0.132	\\	\cline{2-6}
												
                                	&	Intercept	&	$\alpha^0_{CF}$	&	0.168	&	0.096	&	0.079	\\	
Effect of functional  autonomy  	& MCI	&	$\alpha^1_{CF}$		&	-0.030	&	0.097	&	0.760	\\	
                                & dAD 		&	$\alpha^2_{CF}$	&	-0.034	&	0.113	&	0.766	\\	\hline

\multicolumn{3}{l}{\bf{Influence on functional autonomy:}} & & &\\												
                            	&	Intercept	&	$\alpha^0_{FA}$	&	0.114	&	0.045	&	0.012	\\	
Effect of cerebral anatomy 	& MCI 		&	$\alpha^1_{FA}$	&	0.022	&	0.053	&	0.681	\\	
                            & dAD 		&	$\alpha^2_{FA}$	&	-0.053	&	0.069	&	0.441	\\	\cline{2-6}
												
                            &	Intercept	&	$\alpha^0_{FC}$	&	0.111	&	0.049	&	0.023	\\	
Effect of cognitive ability 	& MCI 		&	$\alpha^1_{FC}$	&	0.021	&	0.053	&	0.696	\\	
                            & dAD 		&	$\alpha^2_{FC}$	&	0.082	&	0.072	&	0.259	\\	\cline{2-6}
												
                            &	Intercept	&	$\alpha^0_{FF}$	&	-0.605	&	0.133	&	$<0.001$	\\	
Effect of functional  autonomy 	& MCI 		&	$\alpha^1_{FF}$	&	0.553	&	0.102	&	$<0.001$	\\	
                            & dAD 		&	$\alpha^2_{FF}$	&	0.403	&	0.106	&	$<0.001$	\\	   \hline														
\end{tabular}														
														
\end{table}

\subsubsection{Goodness-of-fit of the model}														
\begin{figure}[p]														
\centering														
\includegraphics[scale=0.5]{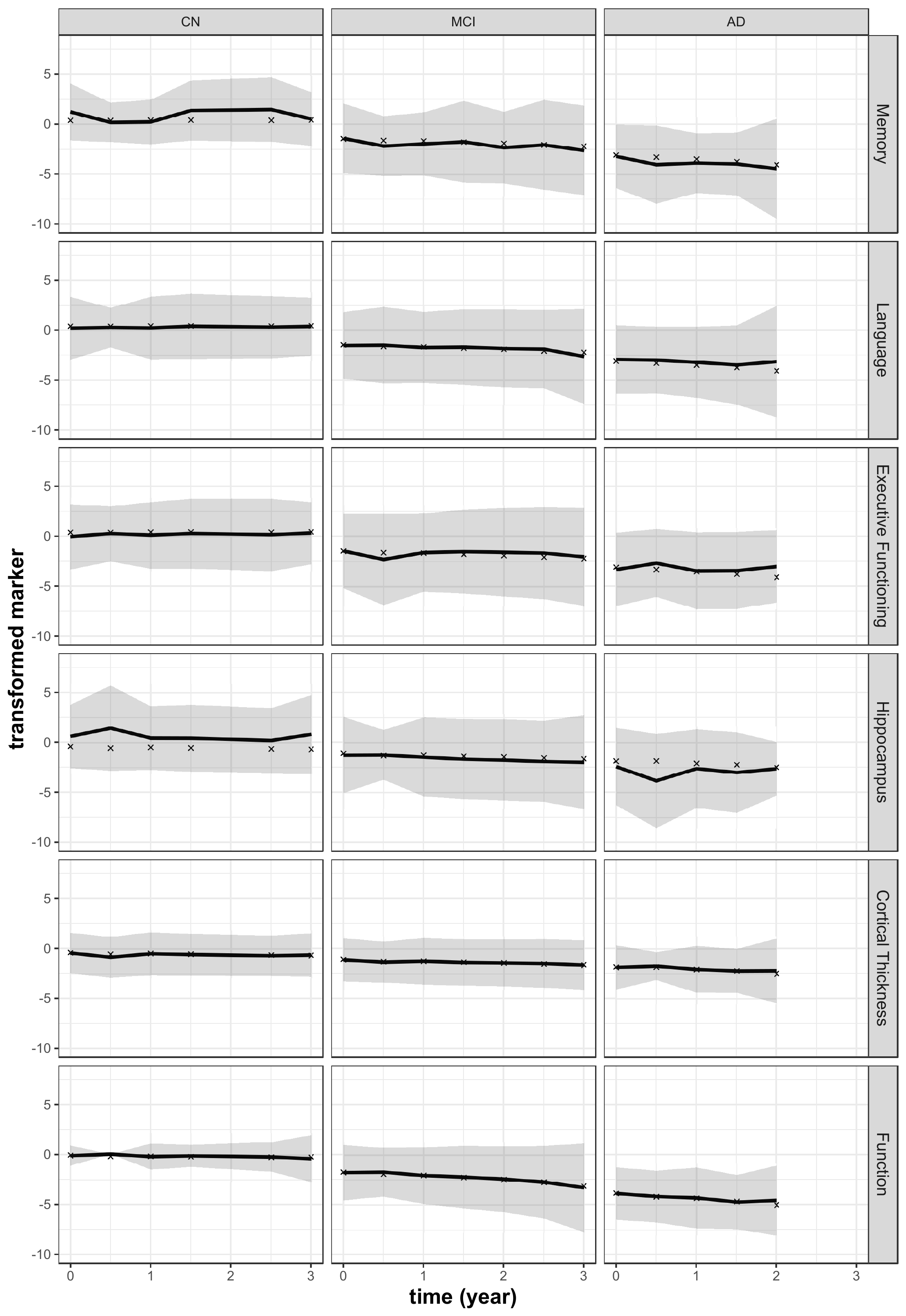}			
\caption{Mean transformed observed scores (plain lines) along with their 95\% confidence interval (shadows) and corresponding mean subject-specific predictions (crosses) by protocol visit. Columns refer to the group (controls CN, subjects with Mild Cognitive Impairments MCI and subjects diagnosed with Alzheimer's disease dAD). Rows refer to the markers (e.g., memory score, language score, executive functioning score, mean hippocampal volume, mean regional cortical thickness and FAQ score) in their transformed scales (see Web Figure 3 for the plot of the estimated link functions between the natural scale and the transformed scale).}
\label{appl_figure3}														
\end{figure}

We assessed the goodness-of-fit of the model by comparing the subject-specific predictions with the observations of the markers in their transformed scale. On the original data, the predictions and observations, summarized by protocol visits and displayed in Figure \ref{appl_figure3}, were very close showing the good fit of the model. %These trajectories also illustrate the differences in trajectories according to clinical stage with each dimension more and more impaired and decline steeper and steeper with disease progression.
 
We also used a 5-fold cross-validation technique to avoid overoptimism in the predictions. Specifically, we split the sample into five groups (each one containing 20\% of the subjects), and estimated the model using data from each combination of 4 groups and computed the subject-specific predictions on the remaining group. The predicted \emph{versus} observed trajectories obtained on the 5 remaining groups, displayed in the Figure 3 of the Web Appendix C, confirmed the good fit of the model to the data.

%---------------------------------------------------------------------														
\section{Discussion}\label{Discussion}														
														
%We proposed in this paper a new statistical model for longitudinal multivariate data which combines a system of difference equations with the mixed %model theory. As such, it has the main advantage to simultaneously describe the trajectories of multiple longitudinal dimensions and assess their %dynamic causal relationships. Until now, multiple longitudinal dimensions have been mostly analyzed with multivariate linear mixed models %\citep{robitaille2012multivariate, mungas2005longitudinal, Fieuws2007}. However, although very useful to assess determinants and shapes of %trajectories, these models do not capture temporal relationships between dimensions but only correlations through correlated random effects. %Dynamic Bayesian Networks and state-space models \citep{liu_mssn_2011, dagum1992dynamic, song2009time} have also been proposed to dissect the %relationships in a network of processes. However, they only relate successive observed states of the network and as such do not apprehend %relationships in a causal, dynamic way \citep{aalen2016can}. Dynamic causal models that rely on a system of differential equations and may %incorporate random effects have been used especially in HIV studies \citep{prague_dynamic_2017} and reach the double objective of describing %precisely a trajectory and assessing causal links. However, their numerical complexity is a major obstacle to their larger spreading to complex %pathologies such as Alzheimer's disease. 

We proposed an original dynamic model to simultaneously describe multivariate processes over time and retrieve temporal relationships between the processes involved. Our model aims to be a dynamic causal model except that we relied on discrete time with difference equations (rather than continuous time with differential equations) to largely reduce the numerical problems, notably with a closed-form likelihood. 

%We fundamentally explored in this work temporal influences between dynamic processes by using the formalism of the dynamic approach to causality notably developed by \cite{aalen2007Whatcan}, \cite{causal_2004_arjas}, \cite{didelez2008}, \cite{commenges_general_2009}. 
%There is also a large literature in statistics on the potential outcomes formalism introduced by \cite{rubin2005causal}; it is however not adapted for \bt{modelling} dynamic relationships between non-directly observed factors, which is the aim in this paper so we do not discuss this approach further.  
This approach goes further in the evaluation of temporal relationships compared to existing approaches such as DBN or CLM which quantify temporal associations by focusing on the effect of a system on the subsequent level of the system at the next visit.
%and they may not easily handle missing or imbalanced data. 
Two major differences are: 
\begin{enumerate}
\item we work at a latent process level to avoid biases in association estimates due to the measurement error and to avoid the necessity to rely on complete and/or balanced observations. The importance of assessing associations at a latent process (or "true" error-free marker) level has been widely documented in a related context: the association between a longitudinal marker and a time-to-event \citep{rizopoulos_joint_2012}.    
\item we assess temporal relationships on the change of processes rather than the state of the system at a subsequent time. We can thus seek local dependence structures in line with the dynamic approach to causality. Yet we acknowledge that causal interpretation still has to be made cautiously as it is always subject to a correct specification of the statistical model.
\end{enumerate}

%\cpl{In contrast,} following the dynamic approach to causality, we chose to analyze the changes of processes  and defined causal relationships in a structural model, that is at a latent process level rather than at an observation level. 

One major simplifying assumption is the time discretization. Fundamentally, causal relationships are to be explored at an infinitesimal level and thus, a causal model is to be defined in continuous time \citep{commenges_general_2009, aalen2016can}. However, in contrast with DBN and CLM discrete approaches, our discretization does not correspond to the visit process; it can be as precise as necessary. It only aims to avoid the numerical complexities due to differential equation modelling and provide a sensible understanding of the rate of change which is modelled. We assessed in a simulation study the impact of the discretization on the underlying time-continuous causal structure; we found that the type-I error rates of the temporal influence parameters were not altered by the use of a model in discrete time, provided the discretization step remained small in regards with the dynamics of the disease under study. Indeed in our simulations and application on Alzheimer's disease, our discretization step was between 1.5 and 6 months while the disease progresses over decades \citep{amieva2008prodromal,jack2013tracking}. In addition to this numerical assessment, we provided in supplementary material analytic approximate relationships between matrices of temporal influences defined in two discretization steps or in continuous and discrete time. %These relationships quantify the associations between temporal influence structures; they remain however approximate as they were found under the assumption that the model structure did not differ substantially between the two discretization steps. This assumption deserves more attention in future work. 
In practice, we recommend to try several reasonable discretization steps, assess the change in goodness-of-fit and stability of the estimates, and retain the most sensible discretization step as a compromise between fit/stability and computational intensity.

Although presented in two parts, our structural model belongs to the family of nonlinear mixed models with specific mean and covariance structures which enable the estimation of temporal associations. Thus our estimation by maximum likelihood benefits from the same properties as mixed models. In particular, it handles imbalanced data and relies on the assumption that missing data (intermittent and monotone) are missing at random. This is realistic in ADNI 1 with its very short follow-up (3 years max). However, in applications with a longer follow-up, and notably in population-based cohorts, it will be probably necessary to account for informative clinical events, such as dementia and death in our context. %This will be possible by extending our model to a joint analysis of times to event but probably at the price of additional numerical complexities. 														
In addition, in studies with a long follow-up, it may be reasonable to assume that the structure of temporal influences evolves with time. Our model can handle this as shown in the simulations.
%where the temporal influence structure was modelled according to time using regression splines. We did not consider however time-dependent temporal influences structures in the application due to the short follow up. Our objective was to contrast the influences structure between clinical stages, and as such we rather included stage-specific intensities of transition.

The model was primarily motivated by the study of the multiple alterations involved in the dementia process in the elderly. 
%Indeed, dementia is characterized by very long and progressive alterations in different cerebral and clinical dimensions. However, a
Although there is a global agreement on the relevant dimensions in AD, their relationships are still poorly understood and a confusion persists between alterations due to normal aging and alterations due to pathological aging leading to dementia. Thanks to the ADNI data which included subjects at different clinical stages of dementia, we were able to exhibit the global structure of dependences between cerebral, cognitive and functional domains in normal aging. 
%characterized as expected by the cerebral structure which explains a part of the cognitive and functional changes, and the functional and cognitive changes which have bidirectional relationships. We also
We found out that the main evolution in this structure due to the pathological process towards dementia (in subjects with mild cognitive impairment or diagnosed with AD) was on cognitive functioning with some effect of cognitive functioning on change in cerebral structure in addition to its effect on change in functional structure. 
%Such backward influence of cognition on anatomic structure may appear counterintuitive from a biological point of view. Yet, some studies showed that interventions made on cognitive functioning (e.g., literacy at adulthood) could induce an improvement in cerebral structure \citep{carreiras2009anatomical, boyke2008training}. 
This application which gives a first insight on the possibilities of this dynamic model, could be now refined by targeting specific brain regions (separating for instance cortical thicknesses from hippocampal volume) and specific cognitive functions (separating for instance memory from executive functioning).

As a conclusion, we proposed here a new methodology that may help identify temporal structures in multivariate longitudinal data. Although applied in dementia context, this approach has potential to address unsolved questions in many other chronic diseases where multiple processes and/or markers are in play. This is the case in other neurodegenerative diseases (e.g., Parkinson, Multi-System Atrophy, Amyotrophic Lateral Sclerosis) but also beyond, for instance in chronic renal disease. The methodology could also help understand how dynamic exposures relate with markers of disease progression and disentangle in particular the direction of temporal associations when reverse causation (i.e., changes in the exposure due to early disease changes) may intervene.

\section*{Acknowledgements}

This work benefited from the support of the project SMALA (ANR-15-CE37-0002) of the French National Research Agency (ANR).
Computer time was provided by the computing facilities MCIA (M\'esocentre de Calcul Intensif Aquitain) at the Universit\'e de Bordeaux and the Universit\'e de Pau et des Pays de l'Adour. Data collection and sharing for this project was funded by the Alzheimer's Disease Neuroimaging Initiative (ADNI) (National Institutes of Health Grant U01 AG024904) and DOD ADNI (Department of Defense award number W81XWH-12-2-0012). 

\bibliographystyle{biom}														
\bibliography{main}

\newpage

\setcounter{section}{0}
\renewcommand{\thesection}{Appendix \Alph{section}:}
%--------------------------------------------------------------------------
\section{Relationships between model components under different discretization steps} 
\label{II-Impact of the discretization}

For computational reasons, the latent processes in the structural model are assumed to evolve in discrete time with a constant step $\delta $ which may vary depending on the application. In reality the processes evolve in continuous time and causal inference should be done in continuous time. We thus established the relationship between transition matrices obtained under different discretization steps: $\delta^*$ and a smaller discretization step $\delta$ with $\delta^* = \rho \times \delta$ and integer $\rho > 1$.				

With a discretization step $\delta = \frac{\delta^*}{\rho}$, the second line of equation (1) in the main paper can be rewritten:
\begin{equation}\label{DiscretizationEffect_1}						
\boldsymbol{\Lambda_i}(t+\delta) =   \delta \left(\boldsymbol{X_i}(t+\delta)\boldsymbol{\gamma}  + \boldsymbol{Z_i} (t+\delta)\boldsymbol{v_i}\right) +  \left(\boldsymbol{I}_D + \delta \boldsymbol{A_{i,\delta}}(t)\right) \boldsymbol{\Lambda_i}(t) ,~ \forall t > 0, \rho > 1 							
\end{equation}	
By recurrence, the network level at $t+ \rho \delta$, that also correspond to $t+\delta^*$, can be expressed as a function of $\boldsymbol{\Lambda_i}(t+ \delta^*)$ as follows:
\begin{equation}\label{DiscretizationEffect_2}			
\boldsymbol{\Lambda_i}(t+ \delta^*) = \boldsymbol{\Psi_{i,\delta}}(t,\rho,0)\boldsymbol{\Lambda_i}(t) +\displaystyle \delta \sum\limits_{s=1}^{\rho}\boldsymbol{\Psi_{i,\delta}}(t,\rho,s)\left( \boldsymbol{X_i}(t+ \delta s) \boldsymbol{\gamma}  + \boldsymbol{Z_i} (t+ \delta s)\boldsymbol{v_i}\right) 							
\end{equation}							
with function $\Psi_{i,\delta}$ defined in the relation (7) of the main paper.						

Considering directly a discretization step of $\delta^*$, the second line of equation (1) of the main paper can also be rewritten							
\begin{equation}\label{DiscretizationEffect_3}							
\boldsymbol{\Lambda_i}(t+\delta^*) =  \delta^* \left(\boldsymbol{X_i}(t+\delta^*)\boldsymbol{\gamma^*}  + \boldsymbol{Z_i}(t+\delta^*)\boldsymbol{v_i^*} \right) + \left( \boldsymbol{I}_D + \delta^* \boldsymbol{A_{i, \delta^*}}(t)\right) \boldsymbol{\Lambda_i}(t) ,~~~~\forall t > 0 ~~~~ \text{and } \delta^* >0.						
\end{equation}

\paragraph{Relationships between transition matrices under different steps}

Considering that the second parts of equations \eqref{DiscretizationEffect_2} and \eqref{DiscretizationEffect_3} are close enough although the model specifications are different, we obtain:
\begin{equation}\label{DiscretizationEffect_4}							
\boldsymbol{I}_D + \delta^* \boldsymbol{A_{i, \delta^*}}(t) \approx \boldsymbol{\Psi_{i,\delta}}(t,\rho,0) = \displaystyle\prod_{l=0}^{\rho-1} \left(\boldsymbol{I}_D + \frac{\delta^*}{\rho} \boldsymbol{A_{i, \delta}}(t + l\frac{\delta^*}{\rho})\right)\\							
\end{equation}							
						
Equation \eqref{DiscretizationEffect_4} can be rewritten to highlight the relationship between the causal structure $\boldsymbol{A_{i, \delta^*}}$  and $\boldsymbol{A_{i, \delta}}$: 					
\begin{equation}\label{DiscretizationEffect_5}							
\boldsymbol{A_{i, \delta^*}}(t)  = \frac{1}{\delta^*} \left(\displaystyle\prod_{l=0}^{\rho-1} \left(\boldsymbol{I}_D + \frac{\delta^*}{\rho} \boldsymbol{A_{i, \delta}}(t + l\frac{\delta^*}{\rho})\right) - \boldsymbol{I}_D \right)							
\end{equation}							
							
When assuming that the causal structure $\boldsymbol{A_{i, \delta}}$ is constant in each interval $\left[t, t+\delta^*\right]$, $\forall t \in \tau$, the relationship \eqref{DiscretizationEffect_5} becomes :							
							
\begin{equation}\label{DiscretizationEffect_6}							
\boldsymbol{A_{i, \delta^*}}(t)  = \frac{1}{\delta^*}\left(\left(\boldsymbol{I}_D + \frac{\delta^*}{\rho} \boldsymbol{A_{i, \delta}}(t)\right)^{\rho} -  \boldsymbol{I}_D \right) 							
\end{equation}							
Or equivalently,
\begin{equation}\label{DiscretizationEffect_7}							
\boldsymbol{A_{i, \delta}}(t)  = \frac{\rho}{(\delta^*)^{(\rho+1)}}\left(\left(\frac{1}{\delta^*}\boldsymbol{I}_D + \boldsymbol{A_{i, \delta^*}}(t)\right)^{\frac{1}{\rho}} -  \boldsymbol{I}_D \right) 				
\end{equation}	

Note that the limits of the right part of relation \eqref{DiscretizationEffect_6}, when $\rho$ tends to infinity provide the relationship between $\boldsymbol{A_{i, \delta^*}}(t)$ and its continuous time counterpart. Assuming that $\lim\limits_{\rho \to \infty} \boldsymbol{A_{i,\delta}}(t) = \boldsymbol{A_i}(t) $ is  finite, the relationship \eqref{DiscretizationEffect_6}, become:	
\begin{equation}\label{DiscretizationEffect_8}							
\boldsymbol{A_{i, \delta^*}}(t) \approx  \frac{1}{\delta^*} \left( \exp{\left(\delta^* \boldsymbol{A_i}(t)\right)} - \mathbf{I}_D \right)\\		\end{equation}							
							
%\begin{equation}\label{DiscretizationEffect_9}							
%A_{i, \delta*}  \approx \frac{1}{\delta^*} \left( e^{\delta^* A_i} - \mathbf{I}_D \right)\\			\end{equation}							

\paragraph{Relationships between trend parameters under different steps}

To relate the transition matrices under different steps, we considered second parts of equations \eqref{DiscretizationEffect_2} and \eqref{DiscretizationEffect_3} were close enough. We identified: 

\begin{equation}\label{equivalenceXZ1}		
\begin{cases}
\boldsymbol{X_i}(t+\delta^*)\boldsymbol{\gamma^*}   \approx \displaystyle \frac{1}{\rho}\sum\limits_{s=1}^{\rho}\boldsymbol{\Psi_{i,\delta}}(t,\rho,s) \boldsymbol{X_i}(t+ \frac{\delta^*}{\rho}s)\boldsymbol{\gamma}  & \\	
\boldsymbol{Z_i} (t+\delta^*)\boldsymbol{v_i^*} \approx \displaystyle \frac{1}{\rho}\sum\limits_{s=1}^{\rho}\boldsymbol{\Psi_{i,\delta}}(t,\rho,s)\boldsymbol{Z_i} (t+ \frac{\delta^*}{\rho}s)\boldsymbol{v_i} & 
\end{cases}
\end{equation}	

In the particular case where  $\boldsymbol{X_i}$ and $\boldsymbol{Z_i}$ reduce to intercepts,

\begin{equation}\label{equivalenceXZ2}		
\begin{cases}
\boldsymbol{\gamma^*}   \approx \displaystyle \frac{1}{\rho}\sum\limits_{s=1}^{\rho}\boldsymbol{\Psi_{i,\delta}}(t,\rho,s) \boldsymbol{\gamma}  & \\	
\boldsymbol{v_i^*} \approx \displaystyle \frac{1}{\rho}\sum\limits_{s=1}^{\rho}\boldsymbol{\Psi_{i,\delta}}(t,\rho,s)\boldsymbol{v_i} & 
\end{cases}
\end{equation}	
Or equivalently,
\begin{equation}\label{equivalenceXZ3}		
\begin{cases}
\boldsymbol{\gamma}   \approx \displaystyle \rho {\left(\sum\limits_{s=1}^{\rho}\boldsymbol{\Psi_{i,\delta}}(t,\rho,s)\right)}^{-1} \boldsymbol{\gamma^*}  & \\	
\boldsymbol{v_i} \approx \displaystyle  \rho {\left(\sum\limits_{s=1}^{\rho}\boldsymbol{\Psi_{i,\delta}}(t,\rho,s)\right)}^{-1} \boldsymbol{v_i^*} & 
\end{cases}
\end{equation}	

%--------------------------------------------------------------
%--------------------------------------------------------------

\section{Additional results in the simulations studies}\label{II-Result_Simus}
\begin{table}[H]
\small
\centering
\caption{Results of the simulations (1000 replicates of samples with 512 subjects) considering two latent processes, each one repeatedly measured by a marker, with linear trajectories and constant covariate-specific causal structure. ASE is the asymptotic standard error, ESE is the empirical standard error and CR is the coverage rate of the 95\% confidence interval.}
\label{II-t1}
  \begin{threeparttable}
\begin{tabular}{rrrrrrrrrrrrr}														
\hline														
& & \multicolumn{5}{c}{without missing values} & & \multicolumn{5}{c}{with missing values$^\star$}\\														
  \cline{3-7}  \cline{9-13} 														
& $\theta$ & $\hat{\theta}$ & bias$^{\dagger}$ & ESE$^{\ddagger}$ & ASE$^{\ddagger}$ & CR(\%) & & $\hat{\theta}$ & bias & ESE$^{\ddagger}$ & ASE$^{\ddagger}$ & CR(\%)\\ 														
 \cline{1-2} \cline{3-7} \cline{9-13}														

  $\beta_1^1$     & -0.268 & -0.270 & 0.8 & 0.070 & 0.073 & 96.2 &&  -0.271 & 1.0 & 0.072 & 0.074 & 95.5 \\ 
  $\beta_2^1$     & -1.695 & -1.704 & 0.5 & 0.113 & 0.111 & 94.3 &&  -1.702 & 0.4 & 0.117 & 0.114 & 94.4 \\ 
  $\beta_1^2$     & 0.057  & 0.059  & 4.0 & 0.082 & 0.085 & 92.3 &&  0.056  & 1.1 & 0.084 & 0.090 & 94.6 \\ 
  $\beta_2^2$     & -1.749 & -1.758 & 0.5 & 0.134 & 0.128 & 93.4 &&  -1.763 & 0.8 & 0.140 & 0.137 & 94.9 \\ 
  $\gamma_0^1$    & 0.042  & 0.043  & 2.4 & 0.018 & 0.017 & 95.0 &&  0.043  & 3.1 & 0.019 & 0.019 & 96.2 \\ 
  $\gamma_1^1$    & -0.033 & -0.033 & 0.1 & 0.016 & 0.017 & 96.3 &&  -0.033 & 0.9 & 0.017 & 0.017 & 95.9 \\ 
  $\gamma_2^1$    & -0.242 & -0.244 & 0.9 & 0.026 & 0.026 & 95.2 &&  -0.245 & 1.3 & 0.028 & 0.029 & 95.4 \\ 
  $\gamma_0^2$    & -0.097 & -0.099 & 1.6 & 0.038 & 0.037 & 95.5 &&  -0.099 & 2.4 & 0.041 & 0.041 & 94.9 \\ 
  $\gamma_1^2$    & -0.014 & -0.014 & 0.7 & 0.018 & 0.019 & 95.5 &&  -0.014 & 0.9 & 0.019 & 0.021 & 94.7 \\ 
  $\gamma_2^2$ 	  & -0.066 & -0.065 & 2.0 & 0.040 & 0.039 & 93.7 &&  -0.064 & 2.5 & 0.042 & 0.042 & 95.0 \\ 
  L(2,1) 		  & 0.333  & 0.334  & 0.3 & 0.054 & 0.054 & 95.2 &&  0.335  & 0.7 & 0.057 & 0.058 & 94.7 \\ 
  L(3,1) 		  & 0.181  & 0.182  & 0.6 & 0.012 & 0.013 & 95.5 &&  0.182  & 0.8 & 0.014 & 0.015 & 96.4 \\ 
  L(4,2)  		  & 0.066  & 0.066  & 0.8 & 0.013 & 0.013 & 95.5 &&  0.067  & 1.2 & 0.015 & 0.014 & 94.8 \\ 
  L(3,3)  		  & 0.145  & 0.140  & 3.7 & 0.041 & 0.010 & 94.3 &&  0.145  & 0.1 & 0.011 & 0.011 & 96.3 \\ 
  L(4,4)  		  & 0.247  & 0.246  & 0.4 & 0.027 & 0.015 & 94.5 &&  0.247  & 0.0 & 0.016 & 0.016 & 94.7 \\ 
  $\alpha_{11}^0$ & -0.230 & -0.231 & 0.6 & 0.016 & 0.016 & 95.1 &&  -0.232 & 0.7 & 0.019 & 0.019 & 95.6 \\ 
  $\alpha_{11}^1$ & 0.099  & 0.100  & 0.8 & 0.013 & 0.013 & 94.8 &&  0.100  & 0.9 & 0.014 & 0.014 & 95.1 \\ 
  $\alpha_{12}^0$ & 0.118  & 0.119  & 0.8 & 0.017 & 0.017 & 95.6 &&  0.119  & 0.9 & 0.019 & 0.020 & 96.1 \\ 
  $\alpha_{12}^1$ & -0.040 & -0.041 & 1.6 & 0.016 & 0.016 & 95.0 &&  -0.041 & 2.4 & 0.018 & 0.019 & 96.2 \\ 
  $\alpha_{21}^0$ & 0.095  & 0.096  & 1.4 & 0.020 & 0.020 & 95.1 &&  0.097  & 1.7 & 0.022 & 0.022 & 94.3 \\ 
  $\alpha_{21}^1$ & 0.043  & 0.042  & 1.6 & 0.022 & 0.023 & 95.2 &&  0.042  & 1.9 & 0.025 & 0.024 & 93.9 \\ 
  $\alpha_{22}^0$ & -0.399 & -0.400 & 0.3 & 0.027 & 0.026 & 93.6 &&  -0.400 & 0.3 & 0.030 & 0.030 & 94.1 \\ 
  $\alpha_{22}^1$ & 0.319  & 0.320  & 0.3 & 0.026 & 0.025 & 94.3 &&  0.320  & 0.4 & 0.028 & 0.028 & 93.5 \\ 
  $\sigma_1$      & 0.397  & 0.399  & 0.5 & 0.014 & 0.014 & 95.2 &&  0.399  & 0.6 & 0.015 & 0.015 & 95.3 \\ 
  $\sigma_2$      & 0.672  & 0.676  & 0.6 & 0.028 & 0.028 & 94.8 &&  0.677  & 0.8 & 0.031 & 0.031 & 95.9 \\ 
  $\eta_{01}$     & 3.793  & 3.796  & 0.1 & 0.123 & 0.121 & 94.3 &&  3.794  & 0.0 & 0.125 & 0.124 & 95.1 \\ 
  $\eta_{11}$     & 1.597  & 1.590  & 0.4 & 0.053 & 0.054 & 93.8 &&  1.590  & 0.4 & 0.055 & 0.056 & 94.8 \\ 
  $\eta_{02}$     & 2.601  & 2.602  & 0.0 & 0.109 & 0.106 & 94.3 &&  2.604  & 0.1 & 0.115 & 0.113 & 95.3 \\ 
  $\eta_{12}$     & 1.226  & 1.220  & 0.5 & 0.048 & 0.048 & 94.0 &&  1.219  & 0.6 & 0.053 & 0.052 & 94.8 \\  
   \hline														
\end{tabular}
\begin{tablenotes}
      \footnotesize 
      \item $^\star$ (15\% missing occasions, 7\% missing outcomes)
      \item $^{\dagger}$ relative bias (\%),
      \item $^{\ddagger}$ ASE is the asymptotic standard error, ESE is the empirical standard error and CR is the coverage rate of the 95\% confidence interval. 
    \end{tablenotes}
\end{threeparttable}
\end{table}

%--------------------------------------------------------------------
\section{Additional results in ADNI application}\label{III-Result_appli}

\begin{table}[H]
\footnotesize
\centering
\caption{Fixed effect estimates on baseline levels and changes of cerebral, cognitive and functional dimensions using a discretization step of 3 month.  \vspace{1mm}} \label{II-t2}

\begin{tabular}{llrrrrrrrrr}
  \hline
  & & \multicolumn{3}{c}{Cerebral dim.} & \multicolumn{3}{c}{Cognitive dim.} & \multicolumn{3}{c}{Functional dim.}\\ \cline{3-11}
 & Parameter & $\theta$ & SE & p-value & $\theta$ & SE & p-value & $\theta$ & SE & p-value \\ \hline
 
 \multicolumn{2}{l}{\underline{At baseline}} & 																  \\
 
  & Age (entery)            & -0.534 & 0.067 & $<0.001$ & -0.404 & 0.070 & $<0.001$ & -0.013 & 0.073 & 0.857    \\    
  & Male			      & -0.183 & 0.081 & 0.025    & -0.207 & 0.099 & 0.037    & -      & -     & -        \\               
  & High. Educ.	          & -0.224 & 0.107 & 0.036    &  0.620 & 0.132 & $<0.001$ & -      & -     & -        \\            
  & APOE   			      & -0.252 & 0.086 & 0.003    & -0.293 & 0.105 & 0.005    & -0.122 & 0.098 & 0.216    \\       
  & MCI  			      & -0.671 & 0.100 & $<0.001$ & -1.693 & 0.137 & $<0.001$ & -1.663 & 0.130 & $<0.001$ \\ 
  & dAD   			      & -1.412 & 0.127 & $<0.001$ & -3.209 & 0.193 & $<0.001$ & -3.770 & 0.194 & $<0.001$ \\

\multicolumn{2}{l}{\underline{On the rate of change}} & 													  \\    
  
 & Intercept  			  & -0.091 & 0.028 & 0.001 &  0.143 & 0.102 & 0.162 & -0.120 & 0.086 & 0.162          \\   
 & MCI 		  			  & -0.068 & 0.035 & 0.049 & -0.589 & 0.186 & 0.002 &  0.009 & 0.123 & 0.940	      \\
 & dAD 		  			  & -0.237 & 0.095 & 0.012 & -1.296 & 0.394 & 0.001 & -0.558 & 0.280 & 0.047	      \\ 
 & Male 		 		  &  0.014 & 0.020 & 0.469 &  0.038 & 0.059 & 0.525 & -      & -     & -              \\      
 & High. Educ. 		 	  & -0.080 & 0.026 & 0.002 &  0.311 & 0.124 & 0.012 & -      & -     & -              \\   
 & APOE 		 		  & -0.100 & 0.021 & $<$0.001 & -0.194 & 0.067 & 0.004 & -0.137 & 0.048 & 0.005          \\                              
 
  \hline
\end{tabular}
\end{table} 

%---------------------------------------------------------------------
\begin{table}[H]
\centering
\caption{Parameter estimates of the Cholesky decomposition of the variance-covariance matrix of overall random effects on baseline levels and changes of cerebral, cognitive and functional dimensions. Elements at 0 or 1 were fixed; standard errors of estimates are indicated in brackets below.\vspace{1mm}}
\label{II-t3}
\begin{tabular}{l|cccccc}
 						 & $u^A$   & $u^C$   & $u^F$  & $v^A$ & $v^C$& $v^F$ \\
   \hline
$u^A$                    & 1 \\
\multirow{2}{*}{$u^C$}   & 0.369   & 1 \\
					     & (0.051) &   \\
\multirow{2}{*}{$u^F$}   & 0.167   & 0.256   & 1 \\
					     & (0.047) & (0.054) &   \\
\multirow{2}{*}{$v^A$}   & 0.140   & 0       & 0      & 0.112\\
						 & (0.032) & 		 & 		  & (0.012)\\
\multirow{2}{*}{$v^C$}   & 0       & 0.562   & 0      & 0      &  0.001\\
						 &         & (0.146) & 		  &        & (0.018)\\
\multirow{2}{*}{$v^F$}   & 0       & 0       & 0.046  & 0      & 0      & 0.419\\
                         &         &         &(0.060) & 	   & 		& (0.041)\\ 
\hline
\end{tabular}
\end{table}

{\small
\begin{longtable}{lcrrr}
\caption{Parameter estimates of the measurement model. Parameters ($\sigma_{Hip.}$,  $\sigma_{zcortex}$, $\sigma_{zmem}$, $\sigma_{zlang}$,$\sigma_{zexec}$,$\sigma_{Faq.0}$) represent the standard deviation of the measurement errors for the 6 markers. Other parameters given by series of six ($\eta_{Hip.0}$ to $\eta_{Hip.5}$; $\eta_{zcortex.0}$ to $\eta_{zcortex.5}$; $\eta_{zmem.0}$ to $\eta_{zmem.5}$;  $\eta_{zlang.0}$ to $\eta_{zlang.5}$; $\eta_{zexec.0}$ to $\eta_{zexec.5}$ and $\eta_{Faq.0}$ to $\eta_{Faq.4}$) correspond to the parameters of the quadratic I-splines link functions used to transform each marker (e.g., Hippocampal volume, Cortical thickness,  Memory score, Language score, Executive functioning score, Functional autonomy score).}
\label{II-t4}\\
\hline
score & Parameter & Estimate & SE & p-value \\ 
\hline
\endfirsthead
\multicolumn{5}{c}%
{\tablename\ \thetable\ -- \textit{Continued from previous page}} \\
\hline
score & Parameter & Estimate & SE & p-value \\ 
\hline
\endhead

\multirow{7}{*}{Hippocampal volume} 
& $\eta_{Hip.0}$      & -8.556 & 0.489 & $<$0.001 \\
& $\eta_{Hip.1}$      &  0.716 & 0.220 & 0.001 \\
& $\eta_{Hip.2}$      &  1.374 & 0.068 & $<$0.001 \\
& $\eta_{Hip.3}$      &  1.350 & 0.061 & $<$0.001 \\
& $\eta_{Hip.4}$      &  1.561 & 0.076 & $<$0.001 \\
& $\eta_{Hip.5}$      & -0.709 & 0.664 & 0.286 \\

& $\sigma_{Hip}$ &  1.906 & 0.082 & $<0.001$   \\ \hline

\multirow{7}{*}{Cortical thickness} 
& $\eta_{zcortex.0}$  & -5.239 & 0.231 & $<$0.001 \\
& $\eta_{zcortex.1}$  &  0.390 & 0.119 & 0.001 \\
& $\eta_{zcortex.2}$  &  0.234 & 0.095 & 0.014 \\
& $\eta_{zcortex.3}$  &  0.780 & 0.046 & $<0.001$ \\
& $\eta_{zcortex.4}$  &  0.539 & 0.066 & $<0.001$ \\
& $\eta_{zcortex.5}$  &  0.223 & 0.170 & 0.190 \\

& $\sigma_{zcortex}$  & 0.253 & 0.009 & $<0.001$  \\ \hline

\multirow{7}{*}{Memory score}
& $\eta_{zmem.0}$     & -7.962 & 0.374 & $<0.001$ \\
& $\eta_{zmem.1}$     &  0.989 & 0.089 & $<0.001$ \\
& $\eta_{zmem.2}$     &  1.093 & 0.065 & $<0.001$ \\
& $\eta_{zmem.3}$     &  0.862 & 0.076 & $<0.001$ \\
& $\eta_{zmem.4}$     &  0.927 & 0.086 & $<0.001$ \\
& $\eta_{zmem.5}$     &  0.905 & 0.111 & $<0.001$ \\

& $\sigma_{zmem}$ & -1.516 & 0.067 & $<0.001$   \\ \hline

\multirow{7}{*}{Language score}
& $\eta_{zlang.0}$    & -8.428 & 0.418 & $<0.001$ \\
& $\eta_{zlang.1}$    &  0.154 & 0.192 & 0.421 \\
& $\eta_{zlang.2}$    & -8.182 & 1.146 & 0.463 \\
& $\eta_{zlang.3}$    &  2.243 & 0.041 & $<0.001$ \\
& $\eta_{zlang.4}$    &  1.827 & 0.064 & $<0.001$ \\
& $\eta_{zlang.5}$    & -1.998 & 0.995 & 0.045 \\

& $\sigma_{zlang}$ &  1.151 & 0.056 & $<0.001$   \\ \hline

\multirow{7}{*}{Executive Functioning score} 
& $\eta_{zexec.0}$    & -8.351 & 0.428 & $<0.001$ \\
& $\eta_{zexec.1}$    &  1.008 & 0.126 & $<0.001$ \\
& $\eta_{zexec.2}$    &  0.568 & 0.130 & $<0.001$ \\
& $\eta_{zexec.3}$    &  1.656 & 0.052 & $<0.001$ \\
& $\eta_{zexec.4}$    &  1.125 & 0.100 & $<0.001$ \\
& $\eta_{zexec.5}$    &  0.467 & 0.333 & 0.160 \\

& $\sigma_{zexec}$ & 1.401 & 0.067 & $<0.001$   \\ \hline

\multirow{7}{*}{Functional autonomy score}
& $\eta_{Faq.0}$      & -7.726 & 0.314 & $<0.001$ \\
& $\eta_{Faq.1}$      &  1.021 & 0.081 & $<0.001$ \\
& $\eta_{Faq.2}$      &  0.269 & 0.140 & 0.054 \\
& $\eta_{Faq.3}$      &  0.565 & 0.064 & $<0.001$ \\
& $\eta_{Faq.4}$      &  0.687 & 0.041 & $<0.001$ \\

& $\sigma_{Faq}$ & 0.657 & 0.027 & $<0.001$   \\ \hline

\end{longtable}
}

%-----------------------------------------
\begin{table}[H]
\centering
\caption{Estimates of the parameters of temporal influences on ADNI data when considering two discretization steps: 3 months (AIC = 29550.5) and 1.5 months (AIC = 29547.1). Superscripts 0, 1 and 2 refer to CN group, MCI group versus CN and dAD group versus CN, respectively.} \label{II-t5}
\begin{tabular}{rrrrrrr}
\hline
 & \multicolumn{3}{c}{$\delta =$ 3 months } & \multicolumn{3}{c}{$\delta =$ 1.5 months }\\ 
\hline
& Estimate & SE & p-value & Estimate & SE  &p-value \\ 
\hline 
$\alpha_{AA}^0$	&	-0.145	&	0.034	&	$<$0.001	&	-0.162	&	0.039	&	$<$0.001	\\
$\alpha_{AA}^1$	&	0.020	&	0.018	&	0.271	&	0.022	&	0.018	&	0.232	\\
$\alpha_{AA}^2$	&	0.002	&	0.025	&	0.946	&	0.002	&	0.027	&	0.932	\\
$\alpha_{AC}^0$	&	0.014	&	0.015	&	0.377	&	0.012	&	0.016	&	0.454	\\
$\alpha_{AC}^1$	&	0.048	&	0.018	&	0.008	&	0.050	&	0.019	&	0.008	\\
$\alpha_{AC}^2$	&	0.058	&	0.027	&	0.030	&	0.063	&	0.028	&	0.025	\\
$\alpha_{AF}^0$	&	0.015	&	0.031	&	0.627	&	0.016	&	0.031	&	0.606	\\
$\alpha_{AF}^1$	&	-0.023	&	0.032	&	0.478	&	-0.024	&	0.032	&	0.447	\\
$\alpha_{AF}^2$	&	-0.040	&	0.037	&	0.286	&	-0.047	&	0.038	&	0.212	\\
$\alpha_{CA}^0$	&	0.267	&	0.078	&	0.001	&	0.299	&	0.089	&	0.001	\\
$\alpha_{CA}^1$	&	0.016	&	0.060	&	0.792	&	0.014	&	0.062	&	0.817	\\
$\alpha_{CA}^2$	&	-0.049	&	0.080	&	0.535	&	-0.056	&	0.085	&	0.511	\\
$\alpha_{CC}^0$	&	-0.548	&	0.148	&	$<$0.001	&	-0.618	&	0.172	&	$<$0.001	\\
$\alpha_{CC}^1$	&	0.171	&	0.055	&	0.002	&	0.175	&	0.060	&	0.004	\\
$\alpha_{CC}^1$	&	0.115	&	0.076	&	0.132	&	0.117	&	0.083	&	0.159	\\
$\alpha_{CF}^0$	&	0.168	&	0.096	&	0.079	&	0.169	&	0.102	&	0.097	\\
$\alpha_{CF}^1$	&	-0.030	&	0.097	&	0.760	&	-0.020	&	0.103	&	0.846	\\
$\alpha_{CF}^2$	&	-0.034	&	0.113	&	0.766	&	-0.033	&	0.121	&	0.787	\\
$\alpha_{FA}^0$	&	0.114	&	0.045	&	0.012	&	0.114	&	0.045	&	0.011	\\
$\alpha_{FA}^1$	&	0.022	&	0.053	&	0.681	&	0.025	&	0.053	&	0.635	\\
$\alpha_{FA}^2$	&	-0.053	&	0.069	&	0.441	&	-0.052	&	0.070	&	0.455	\\
$\alpha_{FC}^0$	&	0.110	&	0.049	&	0.023	&	0.107	&	0.050	&	0.032	\\
$\alpha_{FC}^1$	&	0.021	&	0.053	&	0.696	&	0.030	&	0.054	&	0.584	\\
$\alpha_{FC}^2$	&	0.082	&	0.072	&	0.259	&	0.097	&	0.074	&	0.192	\\
$\alpha_{FF}^0$	&	-0.605	&	0.132	&	$<$0.001	&	-0.555	&	0.130	&	$<$0.001	\\
$\alpha_{FF}^1$	&	0.553	&	0.102	&	$<$0.001	&	0.492	&	0.097	&	$<$0.001	\\
$\alpha_{FF}^2$	&	0.402	&	0.106	&	$<$0.001	&	0.314	&	0.102	&	0.002	\\

\hline
\end{tabular}
\end{table}

%-------------------------------------
\begin{figure}[H]
\centering
\includegraphics[width=1.0\textwidth]{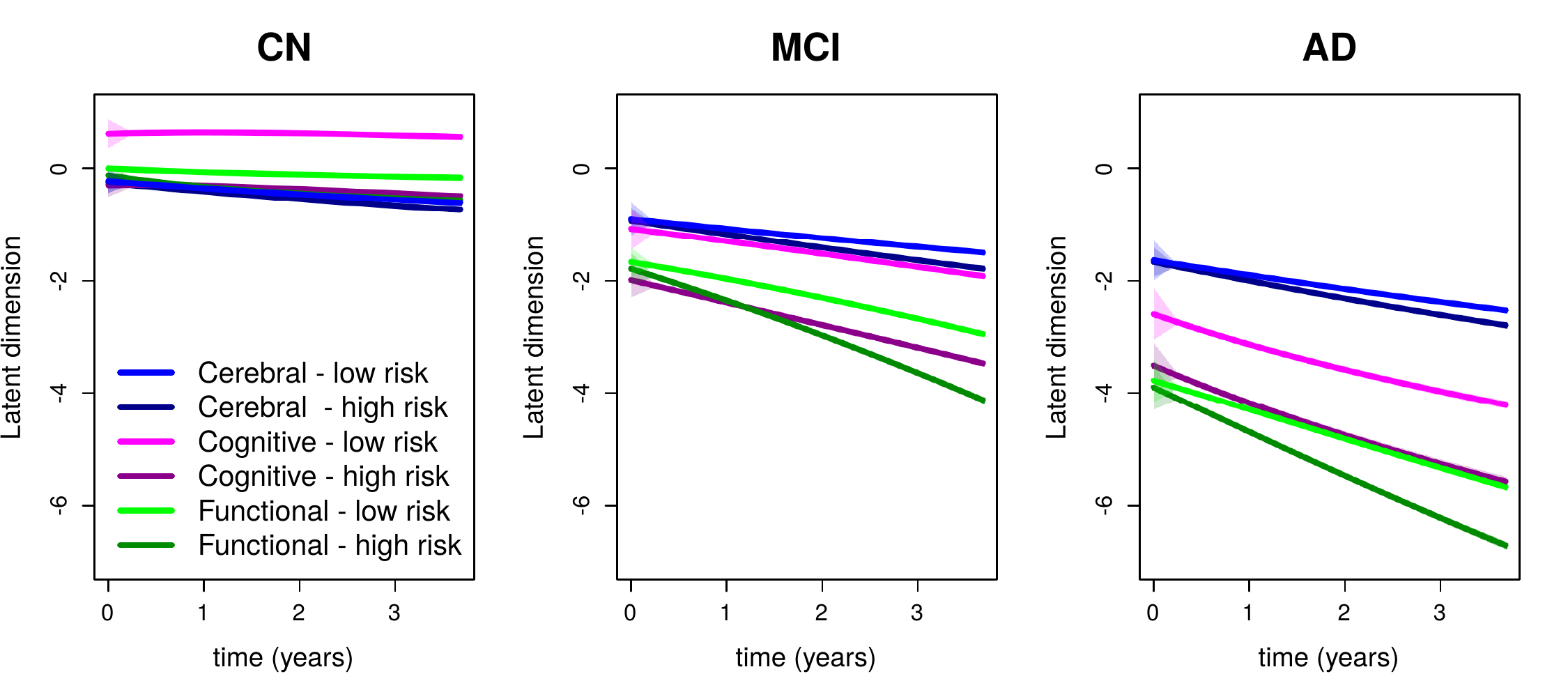}
\caption{Predicted trajectories of cerebral anatomy, cognitive and functional dimensions at healthy (CN), mild cognitive impairment (MCI), and diagnosed with Alzheimer's disease (AD) clinical stages of Alzheimer's disease for a woman with a low risk of Alzheimer's disease (non-carrier of APOE $\epsilon 4$ allele with higher educational level) and for a woman with a high risk of Alzheimer's disease (carrier of APOE $\epsilon 4$ allele with lower educational level).}
\label{II-f1}
\end{figure}

%-------------------------------------
\begin{figure}[H]
\centering
\includegraphics[width=1.0\textwidth]{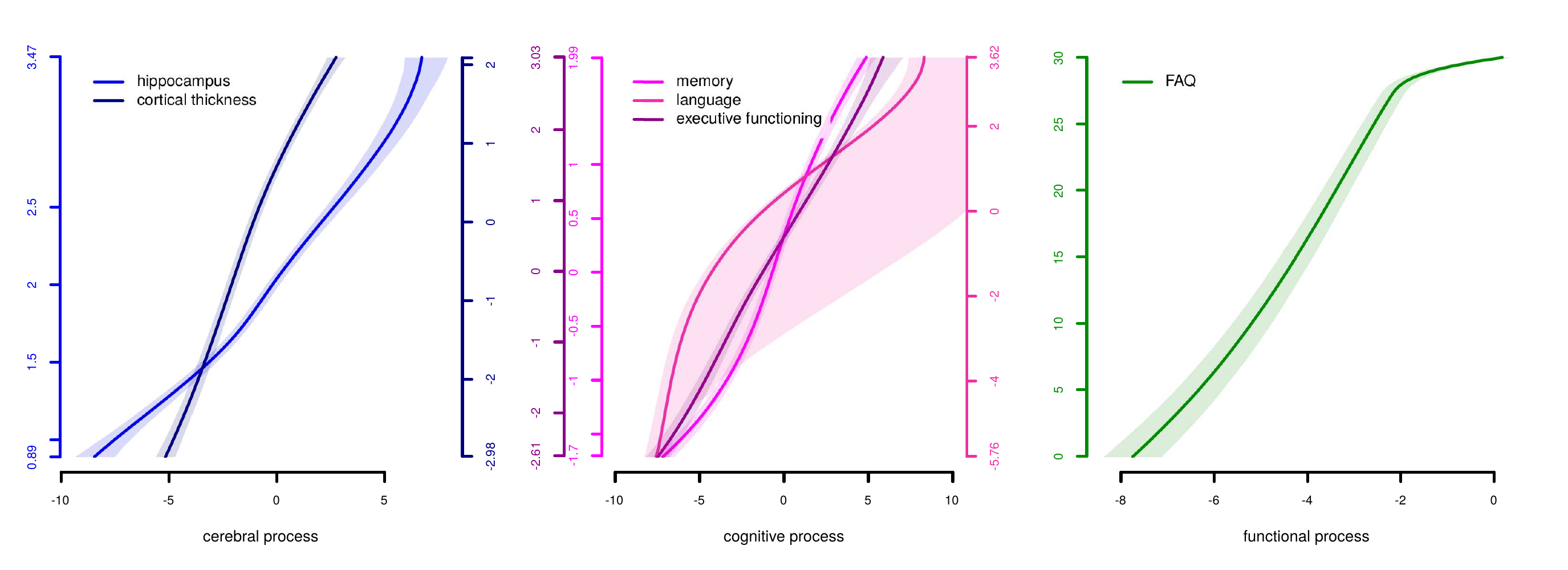}
\caption{Estimated link functions (and 95\% confidence bands obtained by 1000 Monte-Carlo draws) for markers $Y_{\text{hip}}$, $Y_{\text{cortic}}$, $Y_{\text{mem}}$, $Y_{\text{lang}}$,  $Y_{\text{exec}}$ and $Y_{\text{FAQ}}$ (for mean hippocampal volume, mean regional cortical thickness, memory score, language score, executive functioning score and FAQ score).}
\label{II-f2}
\end{figure}
%-------------------------------------

\begin{figure}[H]
\centering														
\includegraphics[scale=0.5]{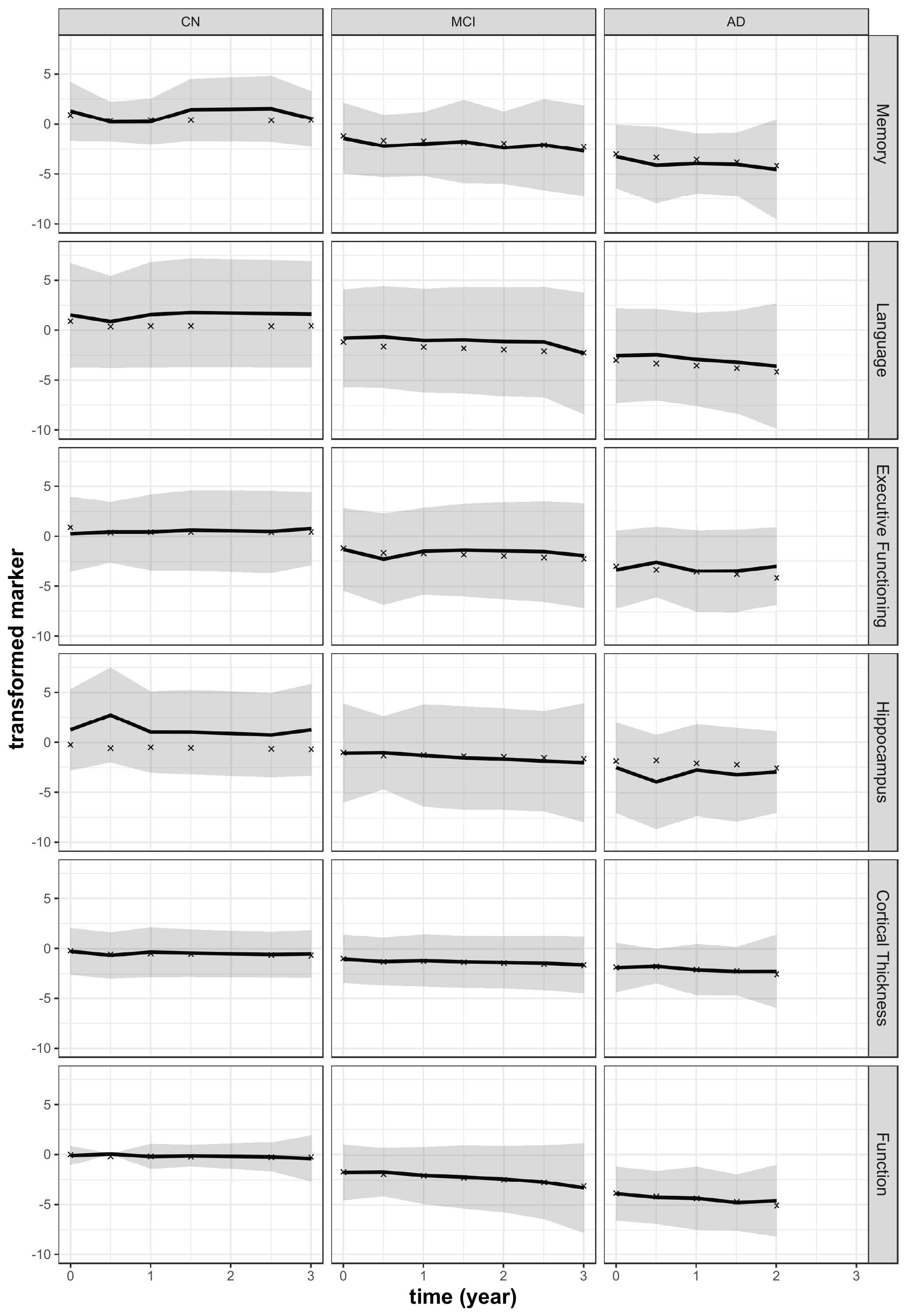}			
\caption{
Mean transformed observed scores (plain lines) along with their 95\% confidence interval (shadows) and corresponding mean subject-specific predictions (crosses) by protocol visit. Predictions are obtained using a 5-fold cross-validation technique. Columns refer to the group (controls CN, subjects with Mild Cognitive Impairments MCI and subjects diagnosed with Alzheimer's disease dAD). Rows refer to the markers (e.g., memory score, language score, executive functioning score, mean hippocampal volume, mean regional cortical thickness and FAQ score) in their transformed scales (see Web Figure 3 for the plot of the estimated link functions between the natural scale and the transformed scale).}													
\label{II-f3}														
\end{figure}

\label{lastpage}

\end{document}